\documentclass[reprint,superscriptaddress,amsmath,amssymb,aps,prb,floatfix]{revtex4-2}
\usepackage{graphicx}% include figure files
\usepackage{dcolumn}% align table columns on decimal point
\usepackage{bm}% bold math
\usepackage[dvipsnames]{xcolor}
\usepackage{braket}
\usepackage{mathrsfs}
\usepackage{appendix}

\begin{document}

\title{Ground state  selection by magnon interactions in the fcc antiferromagnet}
%\title{Order by disorder from magnon interaction in Heisenberg fcc antiferromagnet}

\author{R. Schick}
\affiliation{Institute Laue Langevin, 38045 Grenoble Cedex 9, France}
\affiliation{Univ.\ Grenoble Alpes \& CEA, IRIG, PHELIQS, 38000 Grenoble, France}
\author{O. G\"otze}
\affiliation{Otto-von-Guericke University, Institute of Physics,
P.O.Box 4120, 39016 Magdeburg, Germany}
\author{T. Ziman}
\affiliation{Institute Laue Langevin, 38045 Grenoble Cedex 9, France}
\author{R. Zinke}
\affiliation{ Otto-von-Guericke University,  Institute of Apparatus and Environmental Engineering,
Universit\"atsplatz 2, 39106 Magdeburg, Germany}
\author{J. Richter}
\affiliation{Otto-von-Guericke University, Institute of Physics, 
P.O.Box 4120, 39016 Magdeburg, Germany}
\affiliation{Max-Planck-Institut f\"ur Physik Komplexer Systems, N\"othnitzer Stra\ss e 38, 
01187 Dresden, Germany}
\author{M. E. Zhitomirsky}
\affiliation{Univ.\ Grenoble Alpes \& CEA, IRIG, PHELIQS, 38000 Grenoble, France}

\date{\today}

\begin{abstract}
We study the nearest-neighbor Heisenberg antiferromagnet on a face-centered cubic lattice with 
arbitrary spin $S$. The model exhibits degenerate classical ground states including two collinear
structures  AF1 and AF3 described by different propagation vectors that
are prime candidates for the quantum ground state. We compute the energy for each of the two states as
a function of $S$ using the  spin-wave theory that includes magnon-magnon interaction in a self-consistent way
and the numerical coupled cluster method. Our results unambiguously demonstrate that quantum fluctuations stabilize
the AF1 state for realistic values of spin. Transition to the harmonic spin-wave result, which
predicts the AF3 state, takes place only for $S\agt 10$. We also  study quantum
renormalization of the magnon spectra for both states as a function of spin.
\end{abstract} 

\maketitle

\section{Introduction}
An antiferromagnet on a face-centered cubic lattice has  attracted a longstanding 
theoretical interest \cite{Anderson50,Luttinger51,Li51,Ziman53,Haar62,Lines63,Lines64,Yamamoto72,Swendsen73,
Oguchi85,Henley87,Diep89,Heinilaa93,Yildirim98,Lefmann01,Gvozdikova05,Ignatenko08,Cook15,Batalov16,Sinkovicz16,
Li17,Singh17,Sun18,Balla20,Schick20,Kiese20}.
Early on, Anderson argued for an infinite degeneracy of the classical ground state for the nearest-neighbor
Heisenberg model \cite{Anderson50} making it the second such example after the celebrated
triangular Ising antiferromagnet \cite{Wannier50}. The interest in magnetic frustration on the face-centered cubic (fcc) network
is fueled by an abundance of related materials, see \cite{Seehra88} for survey of the early works and
\cite{Matsuura03,Goodwin07,Aczel16,Chatterji19,Khan19,Revelli19,Bhaskaran21} for more recent studies.

\begin{figure}[tb]
\centering
\includegraphics[width=0.47\columnwidth]{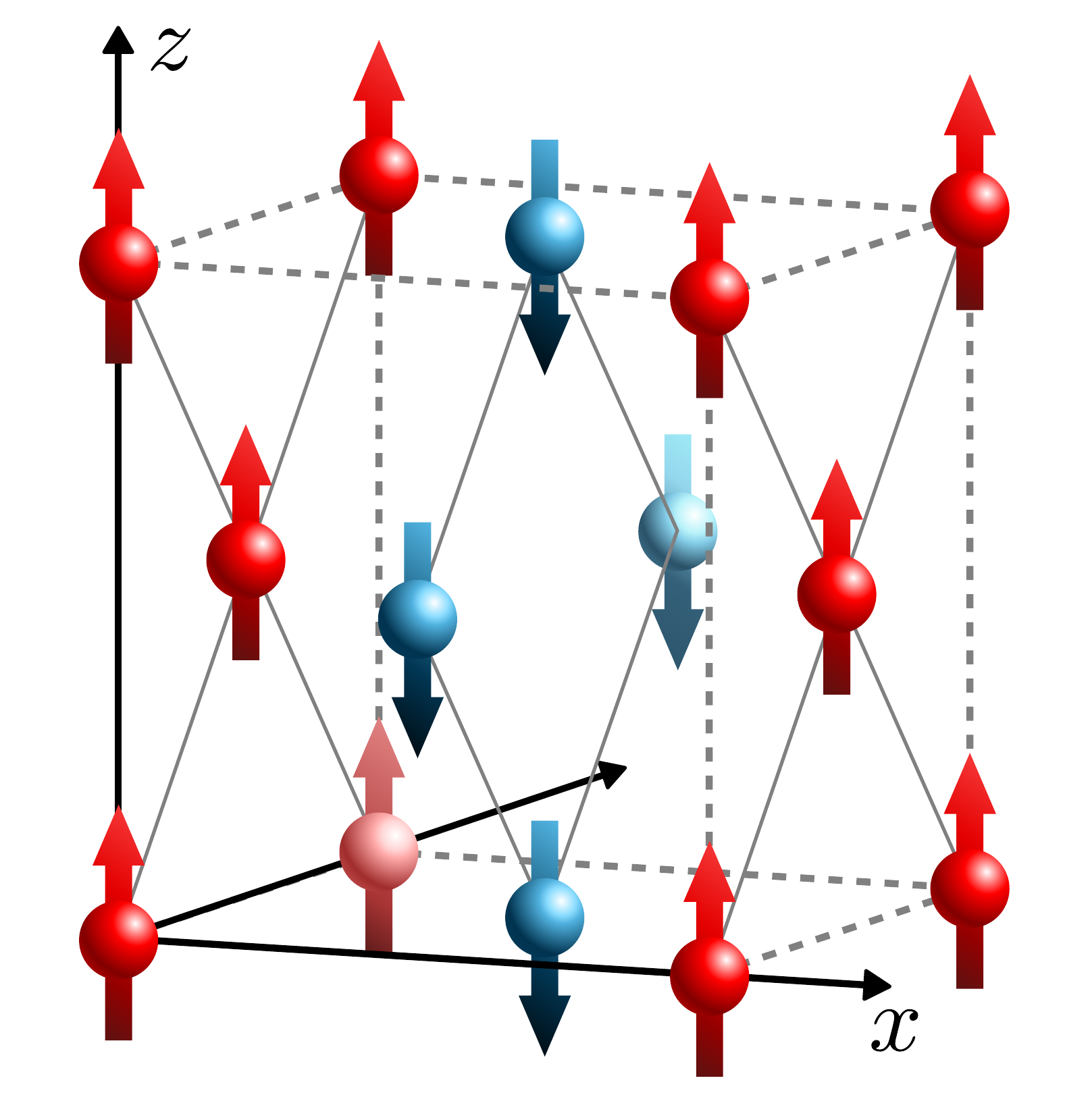}
\hfill
\includegraphics[width=0.47\columnwidth]{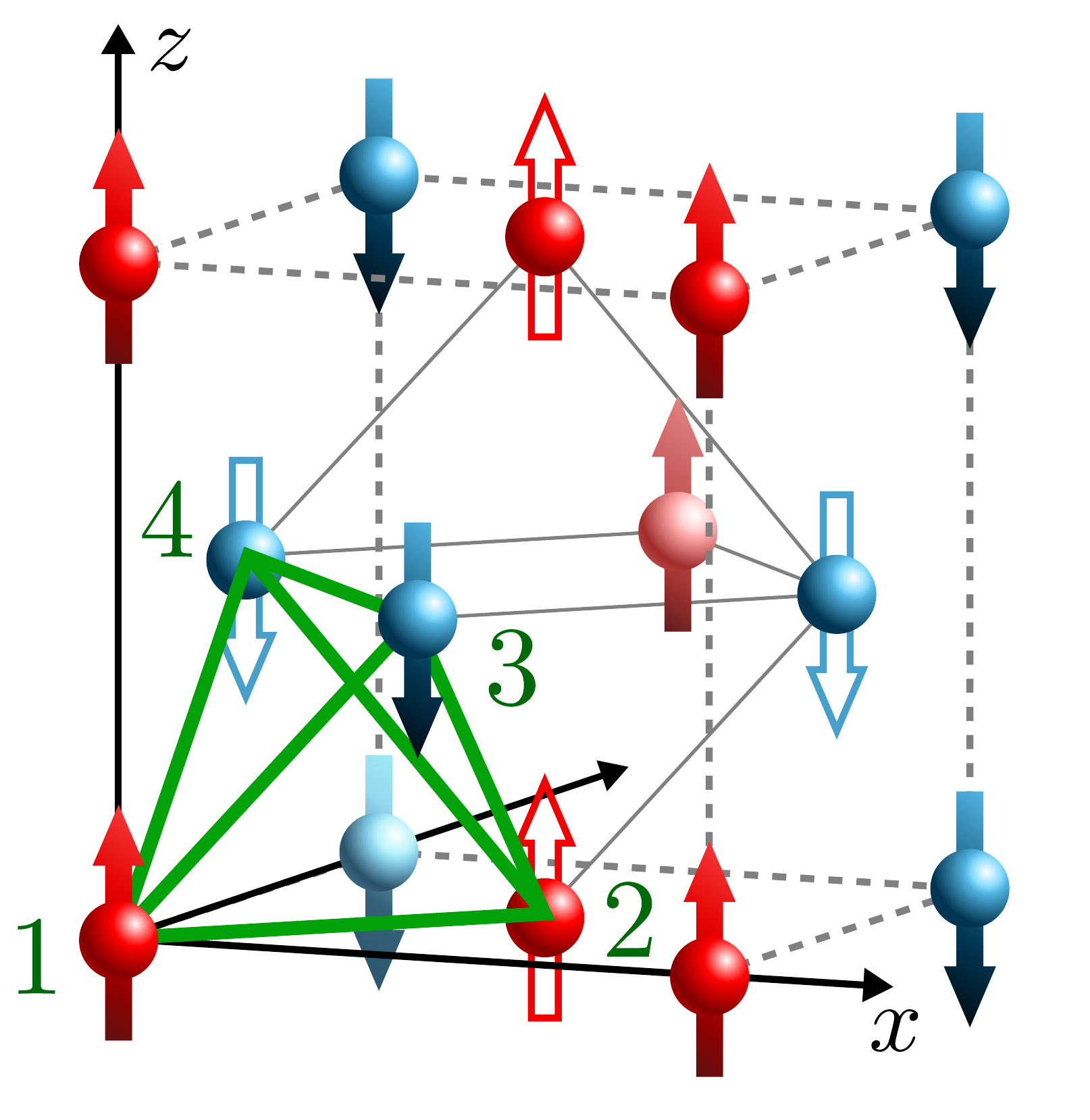}
\caption{Collinear magnetic structures for the nearest-neighbor fcc antiferromagnet.
Left: the two-sublattice AF1 state with ${\bf Q}_1=(2\pi,0,0)$.
Right: the  four-sublattice  AF3  state with ${\bf Q}_3=(2\pi,\pi,0)$.
Closed (open) spins correspond to two rotating sublattices described by $a$ ($b$) bosons
(see the text).
}
\label{fig:AF13}
\end{figure}

The infinite degeneracy in the ground state can be lifted by additional interactions, for example, the
second-neighbor exchange $J_2$ that is often present in the fcc materials  \cite{Seehra88}. The two collinear 
AF1 and AF3 spin structures stabilized, respectively, by weak negative or positive $J_2$ are shown 
in Fig.~\ref{fig:AF13}.  The propagation vector of the AF1 magnetic structure is ${\bf Q}_1 = (2\pi,0,0)$ 
or the two other wavevectors obtained by permutation of its components. The AF3 magnetic structure is described
by ${\bf Q}_3 = (2\pi,\pi,0)$ or other symmetry related vector in the Brillouin Zone (BZ). 
For the nearest-neighbor model ($J_2=0$) the two collinear states become degenerate together with an infinite number of
incommensurate spin spirals described by wavevectors belonging to the line  ${\bf Q}_s = (2\pi,q,0)$ that connects
the AF1 and AF3 wavevectors.

The problem of a finite-temperature transition in an infinitely degenerate frustrated spin model as well 
as subsequent selection of a specific ground state structure by quantum fluctuations was formulated already in the early works
\cite{Li51,Ziman53,Haar62}. The nonzero transition temperature for the nearest-neighbor Heisenberg fcc model
was unambiguously demonstrated by classical Monte Carlo simulations \cite{Diep89,Gvozdikova05}. However, 
the question about its ground state for the quantum model has not been satisfactorily answered. The two main contenders 
are collinear AF1 and AF3 states since quantum effects, often called `order by disorder', usually  
disfavor noncollinear spin arrangements \cite{Shender82,Henley89,Chubukov91,Zhitomirsky15}.
The exact-diagonalization study of the $J_1$--$J_2$ spin-1/2  fcc antiferromagnet in zero field \cite{Lefmann01} 
was performed on clusters up to $N=32$ sites, which is clearly insufficient to reach a conclusion about the type of 
a long-range order for $J_2=0$.
In the recent article, three of us investigated this problem in the harmonic spin-wave
approximation  \cite{Schick20}. The energy difference between the AF1 and  AF3 structures is found to be
\begin{equation}
\Delta E_{13} = E_{\rm AF1} - E_{\rm AF3} = 0.00305(1)JS  \,.
\label{DE13}
\end{equation}
suggesting that the AF3 spin structure is the ground state.
Still, $\Delta E_{13}$ remains very small and the above conclusion may be affected by the  magnon-magnon
interaction. 

The  $1/S$ spin-wave expansion works poorly for highly-frustrated antiferromagnets with lines of pseudo-Goldstone
(zero-energy) modes in the harmonic spectra, see, {\it e.g.}, \cite{Johnston11}.
Instead, in this work we perform a self-consistent spin-wave calculation, which corresponds to summation of an infinite 
sub-series of the $1/S$ diagrams. The magnon interaction renormalizes the bare excitation energies
such that the accidental zero-energy magnons acquire finite quantum gaps.
The ground-state energy correction obtained with the renormalized magnon spectrum is expected to be more reliable.
In addition, we obtain the ground state energies numerically using the coupled-clusters method, 
which appears to be one of a few techniques
suitable for numerical investigation of three-dimensional frustrated magnets. Both approaches agree that the AF1 state is the ground state
of the Heisenberg fcc antiferromagnet for all physical values of spin $S \alt 10$. Furthermore, the absolute energy values
are in good correspondence between the two approaches.

The paper is organized as follows. Section~\ref{sec:SWT} describes the self-consistent spin-wave 
calculations. The general idea of the approach is outlined in Sec.~\ref{sec:SWT}A, whereas analytic results 
for the AF1 and AF3 states are provided in Secs.~\ref{sec:SWT}B and \ref{sec:SWT}C, respectively.
Details of the coupled cluster method (CCM) are presented in Sec.~\ref{sec:CCM}.  
Our main results  for the fcc antiferromagnet obtained by the two methods
are included in Sec.~\ref{sec:OBDO}, where we discuss the ground state properties as well
as quantum renormalization of the magnon spectra. Sec.~V is devoted to 
further discussion and conclusions. Appendices provide extra details  on self-consistent calculations
and spectrum renormalization for the AF3 state, Appen.~\ref{app:AF3}, and on the CCM extrapolation
for $S=1/2$ and 1, 
Appen.~\ref{app:CCM}.
 
% ==============================================================================
% ==============================================================================

%%%%%%%%%%%%%%%%%%%%%%%%%%%%%%%%%%%%%%%%%%%%%%%%%%%%
\section{Spin-wave theory}
\label{sec:SWT}
%%%%%%%%%%%%%%%%%%%%%%%%%%%%%%%%%%%%%%%%%%%%%%%%%%%%
\subsection{Self-consistent approach}

We use the self-consistent spin-wave theory  to study quantum effects in the Heisenberg antiferromagnet 
on an fcc lattice with nearest-neighbor interactions between spins of length $S$:
\begin{equation}
\hat{\cal H} = J\sum_{\langle ij\rangle} {\bf S}_i\cdot{\bf S}_j \,.
\label{H0}
\end{equation}
One of the first formulations of the self-consistent approach was given by Takahashi in a study of
a square-lattice antiferromagnet at finite temperatures \cite{Takahashi89}. 
Various extensions of the Takahashi's work were subsequently applied to ordered and disordered 
quantum magnetic phases at zero and finite temperatures
\cite{Xu90,Barabanov90,Bergomi92,Irkhin92,Gochev94,Dotsenko94,Singh03,Uhrig09,Takano11,Werth18,Yamamoto19}.
We outline details relevant for ordered magnetic states at $T=0$ below.

We use the Holstein-Primakoff representation of spin operators \cite{Holstein40}
\begin{eqnarray} 
&& S^-  =  a^\dagger\,\sqrt{2S-a^\dagger a} \approx \sqrt{2S} a^{\dagger}\Bigl(1 - \frac{a^{\dagger}a}{4S}\Bigr)  , 
\nonumber  \\ 
&& S^+ =  (S^-)^\dagger , \ \ S^z = S - a^{\dagger}a 
\label{HP} 
\end{eqnarray}
applied in the local rotating frame associated with the average spin direction on each site. 
The bond Hamiltonian is, then, expressed via bosonic operators
\begin{equation} 
\hat{\cal H}_{ij}  = J {\bf S}_i\cdot {\bf S}_j \approx E_{ij}^{(0)} + \hat{\cal H}_{ij}^{(2)} + \hat{\cal H}_{ij}^{(4)}  \,,
\label{Hij}
\end{equation} 
restricting expansion up to the fourth order terms.

For the nearest-neighbor fcc antiferromagnet we have to distinguish two types of
bonds with antiparallel ($\uparrow\downarrow$) and parallel ($\uparrow\uparrow$) orientation of 
spins.
In AF1 and AF3  structures  every spin participates in eight $\uparrow\downarrow$ bonds 
and four  $\uparrow\uparrow$ bonds giving them the same classical energy
\begin{equation}
E_0/N  = -2JS^2 \,.
\label{Ecl}
\end{equation}
The quadratic bond contributions are 
\begin{eqnarray}  
\hat{\cal H}_{\uparrow\uparrow}^{(2)} & = & JS
\bigl(a_i^{\dagger}a_j^{_{}} + a_j^{\dagger}a_i^{_{}} - a_i^{\dagger}a_i^{_{}} - a_j^{\dagger}a_j^{_{}}\bigr)\,,  \nonumber \\ 
\hat{\cal H}_{\uparrow\downarrow}^{(2)} & = & JS 
\bigl(a_i^{\dagger}a_i^{_{}} + a_j^{\dagger}a_j^{_{}} - a_ia_j - a_i^{\dagger}a_j^{\dagger}\bigr)\,.  
\label{H_2}
\end{eqnarray} 
The nonlinear quartic terms responsible for magnon-magnon interaction are expressed as
\begin{eqnarray}  
\hat{\cal H}_{\uparrow\uparrow}^{(4)} & = & J \Bigl[n_in_j - \frac{1}{4}
\bigl(a_i^{\dagger}n_i^{_{}} a_j^{_{}} + a_i^{\dagger}n_j^{_{}} a_j^{_{}} + \textrm{h.c.}\bigr)\Bigr]\,, 
 \nonumber \\ 
\hat{\cal H}_{\uparrow\downarrow}^{(4)} & = & J  \Bigl[-n_i n_j + \frac{1}{4} 
\bigl( n_i a_ia_j + n_j a_ia_j  +\textrm{h.c.} \bigr) \Bigr]\,.
 \label{H4} 
\end{eqnarray} 
where h.c.\ stands for the Hermitian conjugate terms and $n_i = a_i^\dagger a_i^{_{}}$ is 
the occupation number operator.

The harmonic or linear spin-wave theory amounts to keeping only quadratic terms in the boson
Hamiltonian. Standard diagonalization of $\hat{\cal H}^{(2)}$ with the help of
the Fourier and the Bogolyubov transformations yields the bare magnon energies. One can also
compute the expectation values of various boson averages in the harmonic ground-state
\begin{equation} 
n = \langle a_i^{\dagger}a_i^{_{}}\rangle \,, \ \ m_{ij} = \langle a_i^{\dagger}a_j^{_{}}\rangle \,, \ \
\Delta_{ij} = \langle a_i a_j \rangle \,.
 \label{MFs}
\end{equation} 
Performing linear spin-wave calculations for a chosen state, it is straightforward to 
verify that the normal averages $m_{ij}$ are nonzero for bonds with parallel spins and vanish
for all antiparallel pairs. The anomalous averages $\Delta_{ij}$ exhibit the opposite pattern: nonzero
for $\uparrow\downarrow$ and zero for $\uparrow\uparrow$ spin pairs.
These relations are imposed by conservation of 
the $z$-component of the total spin for a collinear antiferromagnet with continuous rotation symmetry
about the sublattice direction. Thus, they hold beyond the harmonic approximation in all orders
with respect to magnon-magnon interaction. Once the continuous rotations are absent, either at the level of 
a spin Hamiltonian or because of spin orientation, nonzero $m_{ij}$, $\Delta_{ij}$ appear for every bond.

The next step is to decompose the quartic Hamiltonian  $\hat{\cal H}^{(4)}$ into quadratic terms using 
the standard Hartree-Fock decoupling with the mean-field averages defined by (\ref{MFs}).
Basically, this approximation implies that the magnon scattering process are neglected. Skipping 
straightforward intermediate steps and collecting all relevant contributions
we obtain
\begin{eqnarray} 
\hat{\cal H}_{\uparrow\uparrow} & \approx & J\Bigl[ S^2 - (n - m_{ij})^2 + (S-n+m_{ij})  
\nonumber  \\
&\times&  (a_i^{\dagger}a_j + a_j^{\dagger}a_i - a_i^{\dagger}a_i - a_j^{\dagger}a_j) \Bigr], 
\nonumber \\  
\hat{\cal H}_{\uparrow\downarrow} & \approx & J \Bigl[ -S^2 + \left(n - \Delta_{ij}\right)^2 
\label{Hmf} \\
& + &(S - n + \Delta_{ij}) (a_i^{\dagger}a_i + a_j^{\dagger}a_j - a_ia_j - a_i^{\dagger}a_j^{\dagger}) \Bigr].
\nonumber
\end{eqnarray} 
At this point the mean-field averages are considered as independent parameters and the excitation
spectrum is obtained by diagonalization of a new quadratic Hamiltonian 
$\hat{\cal H}^{(2)}_{\rm MF}$ obtained by summation of bond contributions (\ref{Hmf}).
The system is finally  closed by a self-consistency
condition  (\ref{MFs}), where the averages are computed over a new renormalized ground state.

Once a solution of the self-consistent equations is obtained, the ground state energy can be expressed as
\begin{eqnarray} 
E_{\textrm{g.s.}} & = & E_0 + \langle \hat{\cal H}^{(2)} +  \hat{\cal H}^{(4)}\rangle 
\label{Egs} \\
   & = & J \sum_{\langle ij\rangle}^{\uparrow\uparrow} (S\!-\!n\!+\!m_{ij})^2
 -  J\sum_{\langle ij\rangle}^{\uparrow\downarrow} (S\!-\!n\!+\!\Delta_{ij})^2. 
\nonumber 
\end{eqnarray}
In the following subsections we give explicit analytic results for the AF1 and the AF3 states of the fcc 
antiferromagnet.

\subsection{AF1 state}

The antiferromagnetic AF1 structure consists of two opposite magnetic sublattices which transform into each other 
under translation. As a result, the exchange bonds are characterized by only two mean-field averages:
$m_{ij}=m$  and $\Delta_{ij} = \Delta$ for  parallel and antiparallel pairs of spins, respectively. 
Choosing among three equivalent domains the state with ${\bf Q}_1 = (2\pi,0,0)$, 
we introduce a single species of bosons in the rotating spin frame and obtain for the quadratic part of
the mean-field Hamiltonian (\ref{Hmf}):
\begin{equation}
\hat{\cal H}^{(2)}_{MF} = \sum_{\bf k}\Bigl[ A_{\bf k}a_{\bf k}^{\dagger}a_{\bf k}^{_{}} - \frac{1}{2} B_{\bf k}
\bigl(a_{\bf k}a_{\bf{- k}} + a_{\bf k}^{\dagger} a_{-\bf k}^{\dagger}\bigr)\Bigr] , 
\label{H1_k}
\end{equation}
where 
\begin{eqnarray}
\nonumber
A_{\bf k} & = & 4J(S - n + m)(1 + c_y c_z) + 8J (\Delta - m) \,, 
\nonumber \\
B_{\bf k} & = & 4J(S - n + \Delta) c_x(c_y + c_z) 
\label{AkBk1}
\end{eqnarray}
with $c_\alpha = \cos k_\alpha/2$ for $\alpha = x,y,z$. 
Applying the  Bogolyubov transformation to Eq.~(\ref{H1_k}) one
obtains
\begin{equation}
\epsilon_{\bf k} = \sqrt{A_{\bf k}^2 - B_{\bf k}^2} 
 \label{ek1}
\end{equation}
for the magnon energy, whereas the ground state energy is expressed as
\begin{eqnarray}
 E_{\rm g.s.}/N & =& -2JS^2 + 4J(n-\Delta)^2 - 2J(n-m)^2 
 \nonumber \\
  & + & \frac{1}{2N}\sum_{\bf k}(\epsilon_{\bf k} - A_{\bf k}) \,.
  \label{EgsAF1}
\end{eqnarray}

The self-consistent equations are explicitly given by
\begin{eqnarray}
&& n+\frac{1}{2} = \frac{1}{N}\sum_{\bf k} \frac{A_{\bf k}}{2\epsilon_{\bf k}}\,,\quad
m =  \frac{1}{N}\sum_{\bf k} \frac{A_{\bf k}}{2\epsilon_{\bf k}}\,c_yc_z\,, 
\nonumber \\
&& \Delta  =  \frac{1}{N}\sum_{\bf k} \frac{B_{\bf k}}{4\epsilon_{\bf k}}\,c_x(c_y+c_z)
\label{SC}
\end{eqnarray}
with $\epsilon_{\bf k}$ found from Eqs.~(\ref{AkBk1}) and (\ref{ek1}).
The above equations satisfy the stationary conditions  obtained by varying
the ground state energy (\ref{EgsAF1}) with respect to $n$, $m$, and $\Delta$. Thus, 
the self-consistent solution corresponds to the lowest energy state in the class of variational Bogolyubov
vacuums constructed for the quadratic bosonic Hamiltonians (\ref{Hmf}).

The solution of Eqs.~(\ref{SC}) is found separately for each value of $S$ by iteration procedure starting with 
the harmonic values for $n$, $m$, and $\Delta$. Iterations stop once
an accuracy $10^{-6}$ is reached between two subsequent steps. The final expression for
 the ground state energy of the AF1 state is
\begin{equation}
E_{\textrm{g.s.}}/N = 2J(S - n + m)^2 - 4J(S - n + \Delta)^2 \,.
\label{DE1} 
\end{equation}

\subsection{AF3 state}

The collinear AF3 structure can be represented as
\begin{equation}
{\bf S}_i = \sqrt{2}S\, \hat{\bf z} \cos ({\bf Q}_3\cdot {\bf r}_i  + \pi/4) \,,
\label{AF3struc}
\end{equation} 
where the propagation vector ${\bf Q}_3 = (2\pi,\pi,0)$ or any other
symmetry related vector. The AF3 state has a larger unit cell in comparison to the AF1 structure
with two spins up and two spins down. To simplify analytic calculations we again transform into the rotating local frame.
Still, two type of bosons are needed corresponding to adjacent parallel spins at
$\bm{\rho}_a = (0,0,0)$ and  $\bm{\rho}_b = (\frac{1}{2},\frac{1}{2},0)$, see Fig.~\ref{fig:AF13}.
Within this description, a half of antiparallel pairs $\uparrow\downarrow$  correspond to spins on the same 
rotating sublattice ($a$--$a$ or $b$--$b$)
and the other half is formed by spins from different sublattices ($a$--$b$).
The parallel spin pairs $\uparrow\uparrow$  always belong to different sublattices ($a$--$b$).
Accordingly, in the Hartree-Fock approximation has two independent 
 $\Delta_{ij}$: 
$\Delta_{aa}=\Delta_{bb}$ and $\Delta_{ab}$, whereas $m_{ij}=m$ is unique.
 
After the Fourier transformation, the quadratic boson Hamiltonian can be presented in the matrix form: 
\begin{equation}
\hat{\cal H}^{(2)} = \frac{1}{2}\sum_{\bf k} \Bigl(\hat{X}_{\bf k}^\dagger \,M_{\bf k} \hat{X}_{\bf k}^{_{}} - \Lambda \Bigr) \ , 
\label{H3_k}
\end{equation}
where $\Lambda = \operatorname{Tr}\{M_{\bf k}\}$  and $\hat{X}_{\bf k}^\dagger = 
(a_{\bf k}^{\dagger}, b_{\bf k}^{\dagger}, a_{-\bf k}, b_{-\bf k})$. 
Momentum summation is now performed over the reduced Brillouin zone 
corresponding to the chosen two-sublattice basis, which is shown in the right panel of Fig.~\ref{fig:0modes}. 
The $4\times 4$ matrix $M_{\bf k}$ has the following block structure:
\begin{equation}
 M_{\bf k} = 4J \left(\!
 \begin{array}{cc}
  A_{\bf k} & -B_{\bf k} \\
 -B_{\bf k} &  A_{\bf k}
\end{array}
\!\right)\!, \quad \Lambda = 8J S_0\ ,
\end{equation}
with
the internal  blocks 
\begin{equation}
A_{\bf k}\! =\! \left(\!\begin{array}{cc}
S_0 & S_m\gamma_{\bf k}^* \\ 
S_m\gamma_{\bf k}  & S_0
\end{array}\!\right)\!, \ \
B_{\bf k}\! =\! \left(\!
\begin{array}{cc}
S_a c_xc_z   & S_b \gamma_{\bf k}  \\ 
 S_b \gamma_{\bf k}^* &  S_ac_x c_z 
\end{array}\!\right)\!,
\end{equation}
where $S_0= S-n -m +\Delta_{aa}+\Delta_{ab}$,
$S_m= S-n +m$, $S_a= S-n  +\Delta_{aa}$, $S_b= S-n +\Delta_{ab}$, and
\begin{equation} 
\gamma_{\bf k} =  \frac{1}{2}\,c_y(c_x+c_z) + \frac{i}{2}\,s_y(c_x - c_z)\,,\ \ s_y = \sin\frac{k_y}{2} \,.
\end{equation}

Using the matrix Bogoliubov transformation \cite{White65, Colpa78} 
for diagonalization of the quadratic Hamiltonian (\ref{H3_k}) one obtains
the dynamic matrix
\begin{equation}
 \left| \begin{array}{cc}
 A_{\bf k} - \lambda & -B_{\bf k} \\
B_{\bf k} &  -A_{\bf k}-\lambda
\end{array}
\right| = 0\ ,
\end{equation}
which can be further reduced to 
\begin{equation}
 \left| (A_{\bf k} - B_{\bf k}) (A_{\bf k} +B_{\bf k})-\lambda^2 \right| =0 \ .
 \label{det3}
\end{equation}
Two magnon branches  are given by positive roots of the above biquadratic
equation
\begin{equation}
\epsilon_{\bf k}^\pm = 4J\Bigl[ P_{\bf k} \pm \sqrt{Q_{\bf k}}\Bigr]^{1/2}
\label{ek3}
\end{equation}
with
\begin{eqnarray} 
&& P_{\bf k}   =  S_0^2 - S_a^2 c_x^2c_z^2  
+ \bigl(S_m^2 - S_b^2\bigr) |\gamma_{\bf k}|^2 \,,
\label{PQk} \\
&&Q_{\bf k}   =  4\bigl|S_0S_m\gamma_{\bf k}^*   -   S_aS_b c_xc_z\gamma_{\bf k} \bigr|^2\!
+ S_m^2 S_b^2(\gamma_{\bf k}^2 - \gamma_{\bf k}^{*2})^2   \,.
\nonumber
\end{eqnarray}

In Appendix \ref{app:AF3},
we outline derivation of the self-consistent equations for
$n$, $m$, $\Delta_{aa}$, and $\Delta_{ab}$ without explicitly constructing the Bogolyubov
transformation. Once a solution of self-consistent equations is found,
 the  ground state energy of the AF3 state is expressed as
\begin{eqnarray}
E_{\textrm{g.s.}}/N & = & 2J(S - n + m)^2 - 2J(S - n + \Delta_{aa})^2 
\nonumber \\
& & - 2J(S - n + \Delta_{ab})^2 \ .
 \label{DE3} 
\end{eqnarray}

%%%%%%%%%%%%%%%%%%%%%%%%%%%%%%%%%%%%%%%%%%%%%%%%%%%%%%%%%%%%%%%%%%%%
\section{Coupled cluster method}
\label{sec:CCM}
%%%%%%%%%%%%%%%%%%%%%%%%%%%%%%%%%%%%%%%%%%%%%%%%%%%%%%%%%%%%%%%%%%%%

The coupled cluster method (CCM) has been successfully applied to a variety of quantum frustrated models, see
\cite{zeng98,darradi08,farnell09,kagome_general_s_2011,archi2014,sc-bcc-2015,jiang2015,archi2018,bilayer2019,jian2021}
and references therein. Here  we describe only the basic steps of the CCM calculations.
One starts by choosing a reference quantum state $|\Phi\rangle$, which corresponds usually one of the classical ground states of a 
frustrated spin model.  For the fcc antiferromagnet the two collinear states AF1 and AF3, Fig.~\ref{fig:AF13},
are taken as reference states. Next a rotation to the local frame is performed  such that all spins 
in a reference state align along the negative $z$ axis 
$|{\Phi}\rangle = |\!\downarrow\downarrow\downarrow\ldots\rangle$.
A complete set of multispin creation operators is introduced  in the rotated frame
\begin{equation}
\label{set1} 
C_I^+ = S_i^+ \, , \, S_i^+S_j^+ \, , \, S_i^+S_j^+S_k^+ \, , \, \ldots \ ,
\end{equation}
where
$S^{+}_i = S^{x}_i + iS^{y}_i$, $i,j,k,\ldots$ denote arbitrary lattice
sites, and $C_I^- = (C_I^+)^\dagger$.

The CCM parametrization of bra and ket ground state eigenvectors $\langle\tilde{\Psi}|$  
and $|\Psi\rangle$ of a spin model is chosen as
\begin{eqnarray}
|\Psi\rangle& =& e^{\cal S} |\Phi\rangle \ , \quad {\cal S} =\sum_{I\neq 0}a_IC_I^+ \ ,
\nonumber \\
\langle \tilde{ \Psi}| & = & \langle \Phi |\tilde{\cal S}e^{-{\cal S}} \ , \mbox{ } \tilde{\cal S}=1+
\sum_{I\neq 0}\tilde{a}_IC_I^{-} \; .
\end{eqnarray}
The CCM coefficients $a_I$ and $\tilde{a}_I$ contained  in the correlation operators $\cal S$ and $\tilde{\cal S}$
are determined by %the corresponding  ket-state and bra-state equations
\begin{equation}
\langle\Phi|C_I^-e^{-S}\hat{\cal H}e^S|\Phi\rangle = 0,\ \
\langle\Phi|{\tilde S}e^{-S}[\hat{\cal H}, C_I^+]e^S|\Phi\rangle = 0.
\end{equation}
Each ket- and bra-state equation labeled by a multi-spin index $I$
corresponds to a certain configuration of lattice sites $i,j,k,\ldots$
Using the Schr\"odinger equation, $\hat{\cal H}|\Psi\rangle=E|\Psi\rangle$, one can  write
the ground state energy and the sublattice magnetization as
\begin{equation}
E_{\rm g.s}=\langle\Phi|e^{-\cal S}\hat{\cal H} e^{\cal S}|\Phi\rangle \,,\quad
M = -\frac{1}{N} \sum_i\langle\tilde\Psi|S_i^z|\Psi\rangle\,,
\end{equation}
where $S_i^z$ is computed  in the rotated frame.

In order to truncate the series for $\cal S$ and $\tilde{\cal S}$  we use a standard  
SUB$n$-$n$ approximation scheme \cite{zeng98,kagome_general_s_2011,archi2018}. 
In the SUB$n$-$n$ scheme we include no more than $n$ spin flips spanning a range of no more than
$n$ adjacent lattice sites \cite{SUBn-n}.
This scheme allows us to improve the approximation level in a systemic and controlled manner. 
Using an efficient parallelized code \cite{cccm}, we solved the CCM equations up to the SUB8-8 level 
for $S=1/2$ with $N_c=410 750$
($N_c=1 643 726$) non-equivalent multispin configurations for the AF1 (AF3) state.
%Basically, $N_c$ is dimensionality of an effective Hilbert space used in the CCM calculations
%for given  $n$ and $S$.
For $S>1/2$  multiple on-site spin flips are allowed producing a fast growth of $N_c$ with increasing $S$. 
The highest approximation level is only SUB6-6 except of the AF1 state with $S=1$, for which
we computed the SUB8-8 result  with $N_c= 5 490 340$. On the other hand, 
for the AF3 state with $S=1$ we could not get the SUB8-8 data, because the number of   
non-equivalent multispin configurations is significantly larger 
($N_c= 22 089 437$).

The obtained SUB$n$-$n$ results have to be extrapolated to the $n\to\infty$ limit.
As was established previously \cite{farnell09,archi2014,sc-bcc-2015,archi2018}, 
the extrapolation scheme takes different forms for the ground state energy and the sublattice  
magnetization:
\begin{equation}
E_{\rm g.s.}(n) = a_0 + \frac{a_1}{n^2} + \frac{a_2}{n^4}\,, \ \
M(n)=b_0 + \frac{b_1}{n} + \frac{b_2}{n^2}\,.
\label{SUBextra}
\end{equation}
For $S=1/2$ we can use four data points  $n=2,4,6,8$ as well as their subsets  $n=2,4,6$ and  $n=4,6,8$ 
to check the accuracy of three-parameter fits (\ref{SUBextra}).
However,  for $S=1$, AF3, as well as for  $S>1$  we have only three data points (SUB$n$-$n$, $n = 2, 4, 6$) 
and, thus, performed only a 
single extrapolation. Hence, the obtained CCM results   
 in these cases are generally less accurate.  The CCM results included into the plots of  Sec.~\ref{sec:OBDO}
correspond to extrapolation of the restricted series $n = 2$--6 for all spin values. Further details on the extrapolation results 
for $S=1/2$ and $S=1$ are provided in Appendix~\ref{app:CCM}.

%%%%%%%%%%%%%%%%%%%%%%%%%%%%%%%%%%%%%%%%%%%%%%%%%%%%%%%%%%%%%%%%%%%%
\section{Quantum order by disorder}
\label{sec:OBDO}
%%%%%%%%%%%%%%%%%%%%%%%%%%%%%%%%%%%%%%%%%%%%%%%%%%%%%%%%%%%%%%%%%%%%

% ==============================================================================
\begin{figure}[tb]
\centering
\includegraphics[width = 0.9\columnwidth]{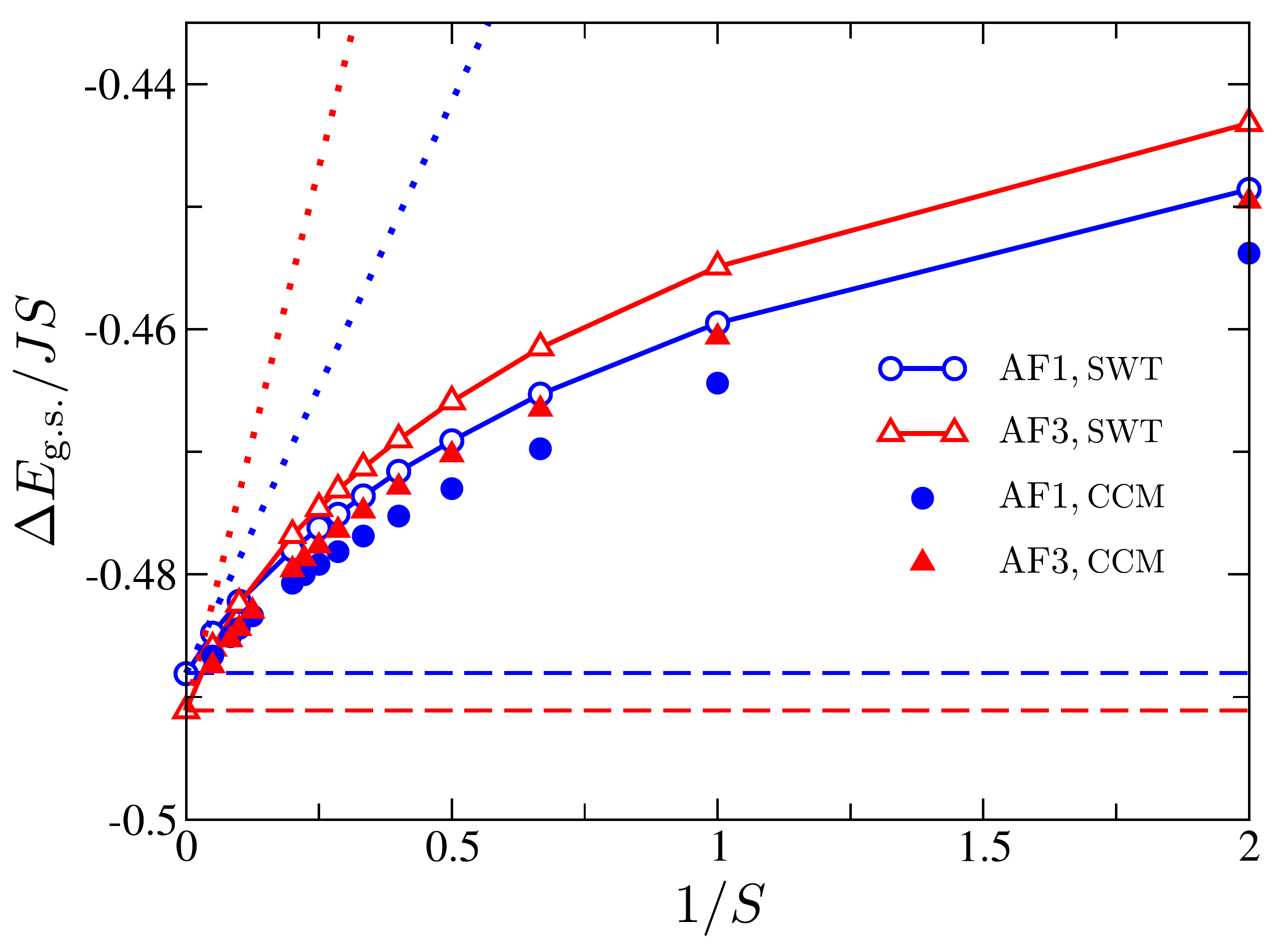} 
\caption{Quantum correction to the ground state energy for the AF1 and AF3 states as 
a function of inverse spin. Continuous lines with open circles and  triangles  are
the self-consistent spin-wave results. Solid symbols indicate the CCM results.
Dashed and dotted lines show the harmonic ($1/S$) and the $(1/S)^2$  
spin-wave energies. 
}
\label{fig:Egs}
\end{figure}
% ==============================================================================

\subsection{Ground-state properties} 

We begin with results for the ground-state energies of the two competing antiferromagnetic structures.
The harmonic spin-wave theory was used in Ref.\ \onlinecite{Schick20} to compute the 
 $1/S$ energy correction.
An earlier work  \cite{Haar62} employed an incorrect magnon spectrum for the collinear 
AF3 state thus coming to an erroneous conclusion, see \cite{Schick20} for further details.
 The next order $(1/S)^2$ energy correction is straightforwardly obtained
using Eq.~(\ref{Egs})  with the harmonic values for
$n$, $m_{ij}$, and $\Delta_{ij}$. The first two terms in the $1/S$ series for 
the ground-state  energies of two states are 
\begin{eqnarray}
E_{\rm g.s.}^{\rm AF1} & = & -2JS^2\Bigl[ 1 + \frac{0.488056}{2S} - \frac{0.186629}{(2S)^2} \Bigr] , 
\nonumber \\
E_{\rm g.s.}^{\rm AF3} & = & -2JS^2\Bigl[ 1 + \frac{0.491106}{2S} - \frac{0.354197}{(2S)^2} \Bigr] .
\label{Egs_per}
\end{eqnarray}
The first-order $1/S$ correction lowers the ground-state energy with respect to the classical value
$E_0=-2JS^2$. The corresponding energy shift is larger for the AF3 state but by a very small 
amount $\Delta E \simeq 0.003JS$.

For frustrated spin models with degenerate classical ground states, an energy gain due to the quantum order by disorder mechanism  
is typically an order of magnitude larger $0.1$--$0.01JS$. Thus, the harmonic zero-point energies of two collinear magnetic structures in the fcc antiferromagnet appear to be accidentally close to each other. In such a case, higher-order quantum corrections resulting 
from magnon-magnon interaction can play a decisive role. For the fcc antiferromagnet, the magnon repulsion yields a state-dependent 
upward shift of the ground-state energies (\ref{Egs_per}).  As a result the net energy gain for the AF1 state appears to be larger than for the AF3 structure modifying the conclusion based on the harmonic theory. 

The convergence and accuracy of the $1/S$ series are, however,
questionable for a frustrated spin model with lines of pseudo-Goldstone (zero-energy) modes. Indeed, the second-order $1/S$ correction to the sublattice magnetization,  $\Delta S \simeq \sum_{\bf k} 1/\epsilon_{\bf k}^3$, diverges for the nearest-neighbor fcc antiferromagnet.      
To overcome the above problem,
we resort to the self-consistent spin-wave calculations described in Sec.~\ref{sec:SWT}.
The renormalized magnon spectrum has only true Goldstone modes and, thus provides
a better starting point for computing various physical properties. 

The effect of quantum renormalization can be illustrated by comparing bosonic averages
in the  AF1 ground state for $S=1/2$ obtained self-consistently
\begin{equation} 
n = 0.14094 \,, \ \ m = 0.07066   \,, \ \ \Delta = 0.16381 
 \label{MF1}
\end{equation}
and from the harmonic spin-wave theory:
\begin{equation} 
n = 0.33878 \,, \ \ m = 0.10994  \,, \ \ \Delta =0.28537 \,.
 \label{MF10}
\end{equation}
The interacting spin-wave vacuum is significantly modified in comparison to the noninteracting ground state.
In particular,  the harmonic theory overestimates $n$ and $\Delta$ by a factor of two. Corresponding 
values for the AF3 structure are presented in Appendix~\ref{app:AF3}.

Figure~\ref{fig:Egs} shows the quantum correction to the classical ground-state energy 
$$
\Delta E_{\rm g.s.} = E_{\textrm{g.s.}} - E_0 
$$ 
normalized to $JS$ and plotted as a function of $1/S$. The full lines with open symbols, circles (AF1)
and triangles (AF3), indicate energies obtained by the self-consistent spin-wave theory.
Numerical CCM results  for both states are  shown by solid symbols. In addition, the  dashed and  
dotted lines indicate energies calculated to the $1/S$ and the $(1/S)^2$ order, respectively.

The total ground-state energies of two collinear states obtained self-consistently for $S=1/2$
\begin{equation}
E_{\rm g.s.}^{\rm AF1} =  -0.72425J,\ \ E_{\rm g.s.}^{\rm AF3} =  -0.72160J 
\end{equation}
differ significantly from  the first- and the second-order spin-wave results (\ref{Egs_per}).
On the other hand, the self-consistent  theory and the CCM  give remarkably consistent values for
 $\Delta E_{\rm g.s.}$  as a function of spin. In particular, the CCM ground-state energies 
 in the case of  $S=1/2$ are
 \begin{equation}
E_{\rm g.s.}^{\rm AF1} =  -0.7267(3)J,\ \ E_{\rm g.s.}^{\rm AF3} =  -0.7244(3)J, 
\label{Egs_CCM}
\end{equation}
 see Appendix \ref{app:CCM} for further details.
 
The AF1 state has a lower energy than the AF3 state for all realistic spin values
 $S<S^* \approx 10$. The harmonic-theory prediction is recovered only 
 for unphysically large spins. The remaining difference between spin-wave values
 and the extrapolated CCM data should be attributed to the magnon scattering processes
that are not included in the self-consistent theory.
 
 % ==============================================================================
\begin{figure}[tb]
\centering
\includegraphics[width = 0.85\columnwidth]{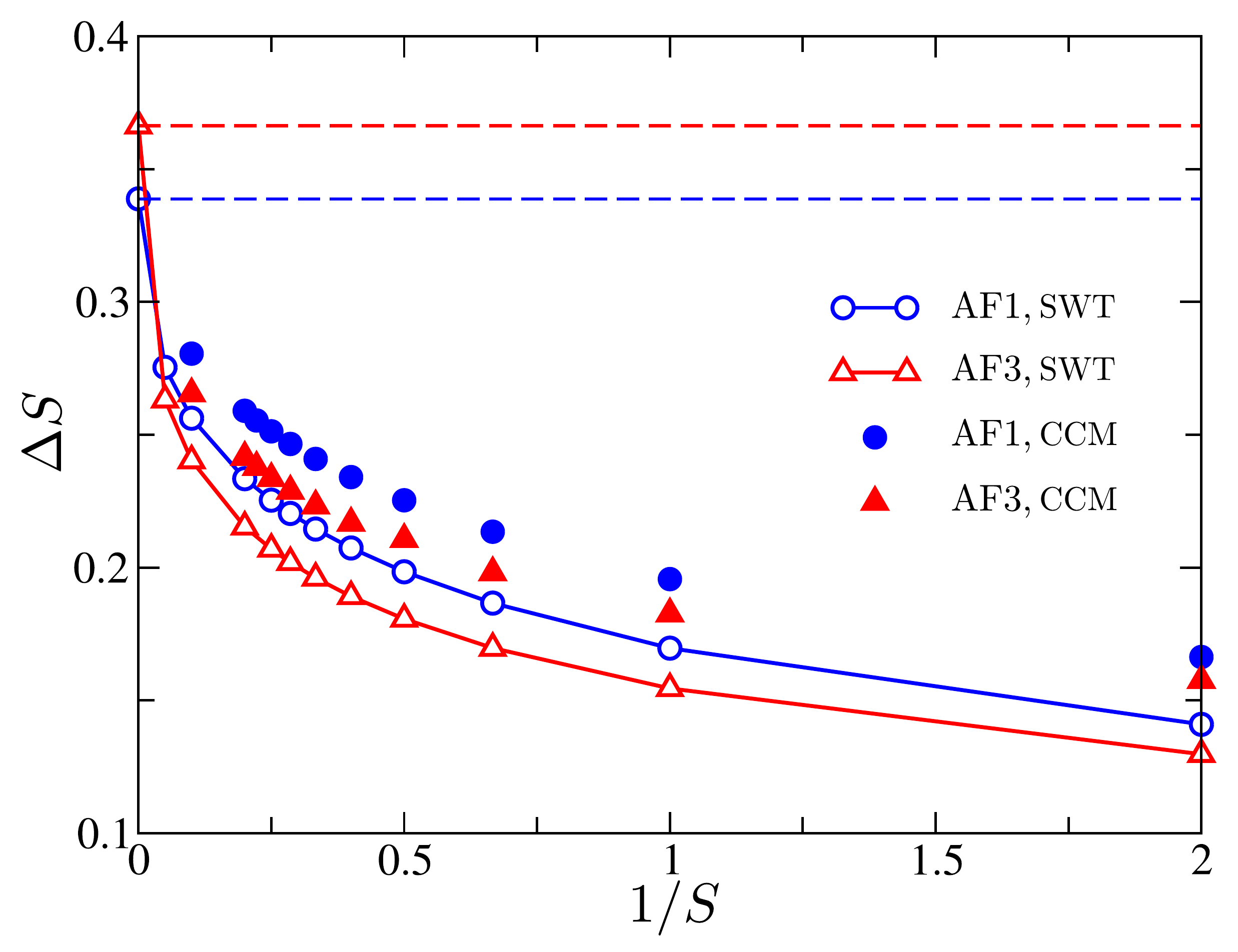}
\caption{Quantum correction to the sublattice magnetization $\Delta S = S-\langle S\rangle$
in two collinear antiferromagnetic structures. Continuous lines with open circles and 
 triangles are obtained in the self-consistent spin-wave approximation. Solid 
symbols represent the CCM results. The horizontal dashed lines indicate the harmonic 
values.
}
\label{fig:Sav}
\end{figure}
% ==============================================================================

Another effect produced by zero-point  fluctuations is reduction of the sublattice magnetization 
from its classical value $\langle S^z\rangle=S$.
Figure~\ref{fig:Sav} shows the spin reduction $\Delta S = S- \langle S^z\rangle$
obtained in the self-consistent spin-wave approximation and from the extrapolation of the CCM results.
The two approaches consistently predict $\Delta S$ to be substantially  smaller than the values 
obtained from the harmonic spin-wave theory. A lack of accuracy of the harmonic approximation 
can be again related to the presence of spurious pseudo-Goldstone modes in the harmonic magnon spectra,
whereas the higher-order quantum corrections restore the correct form of $\epsilon_{\bf k}$, see 
Sec.~\ref{sec:OBDO}B below. For $S=1/2$, the ordered moments in the two antiferromagnetic states
are reduced by about 30\%, which is quite large for a three-dimensional antiferromagnet, but still much 
smaller than the 70\%\ reduction predicted by the harmonic theory \cite{Schick20}.

\subsection{Spectrum renormalization} 

Quantum fluctuations have a profound effect on the excitation spectra of frustrated magnets with classical 
ground-state degeneracy. At the harmonic level, degrees of freedom that connect different ground states show up as
pseudo-Goldstone modes at zero energy. They are shifted to finite energies by quantum corrections, see, for
example, \cite{Belorizky80,Shender82,Chubukov92,Rau18}. Below we discuss such renormalization effects 
focusing on the AF1 state, which is the ground state of the fcc antiferromagnet for all realistic values of spin. 
Complementary results for the AF3 state are presented in Appendix~\ref{app:AF3}.

\begin{figure}[tb]
\centerline{
\includegraphics[width=0.4\columnwidth]{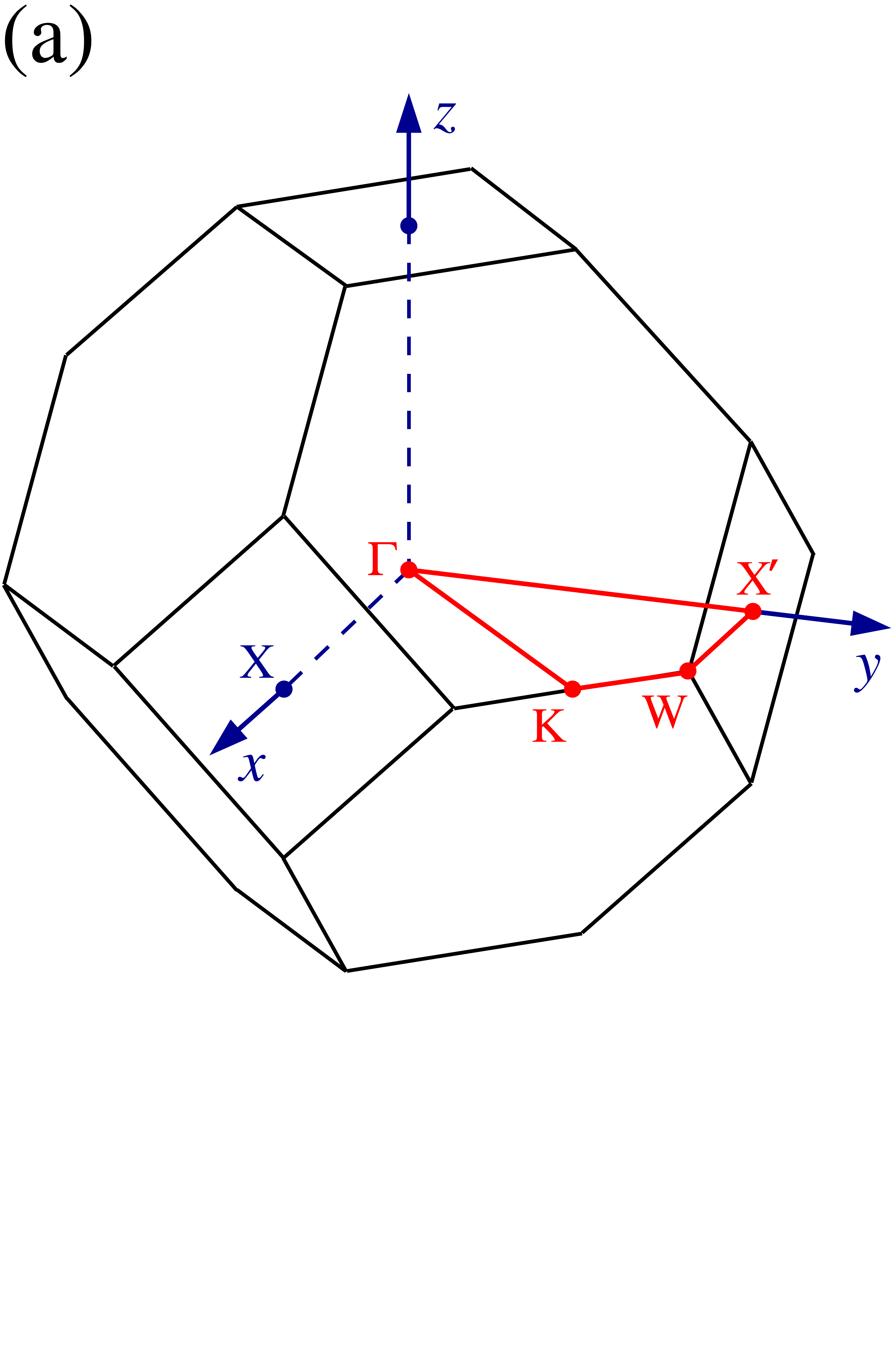}
\hskip 1mm
\includegraphics[width=0.5\columnwidth]{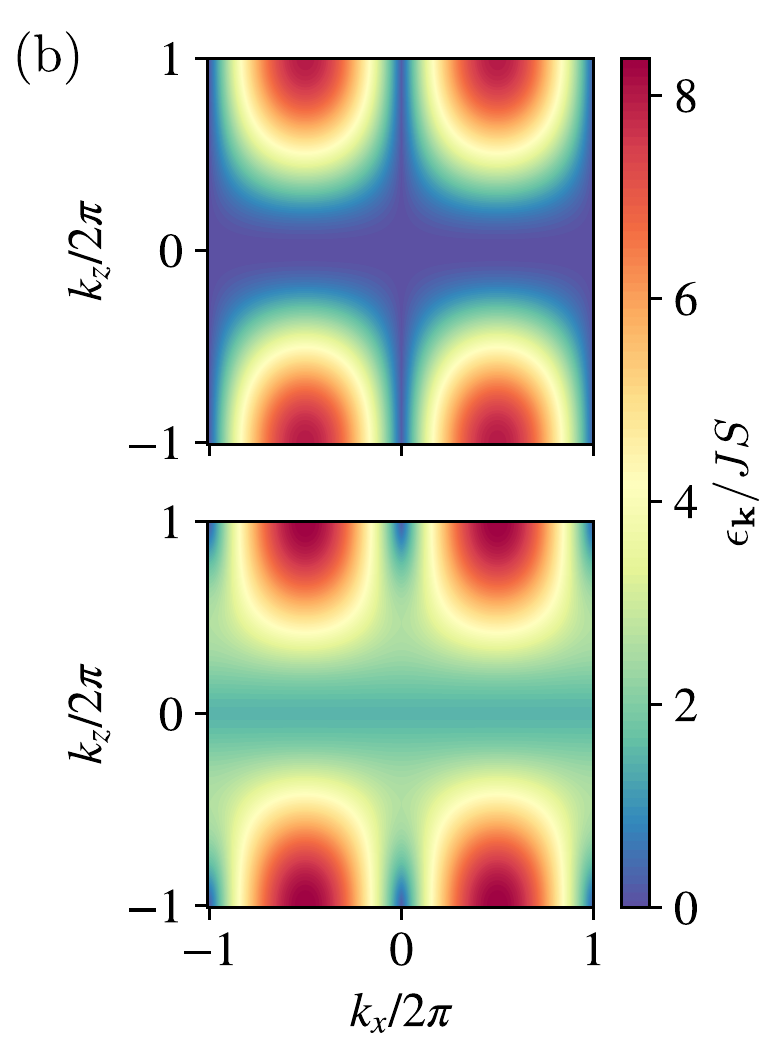}}
\vskip 2mm
\includegraphics[width=0.8\columnwidth]{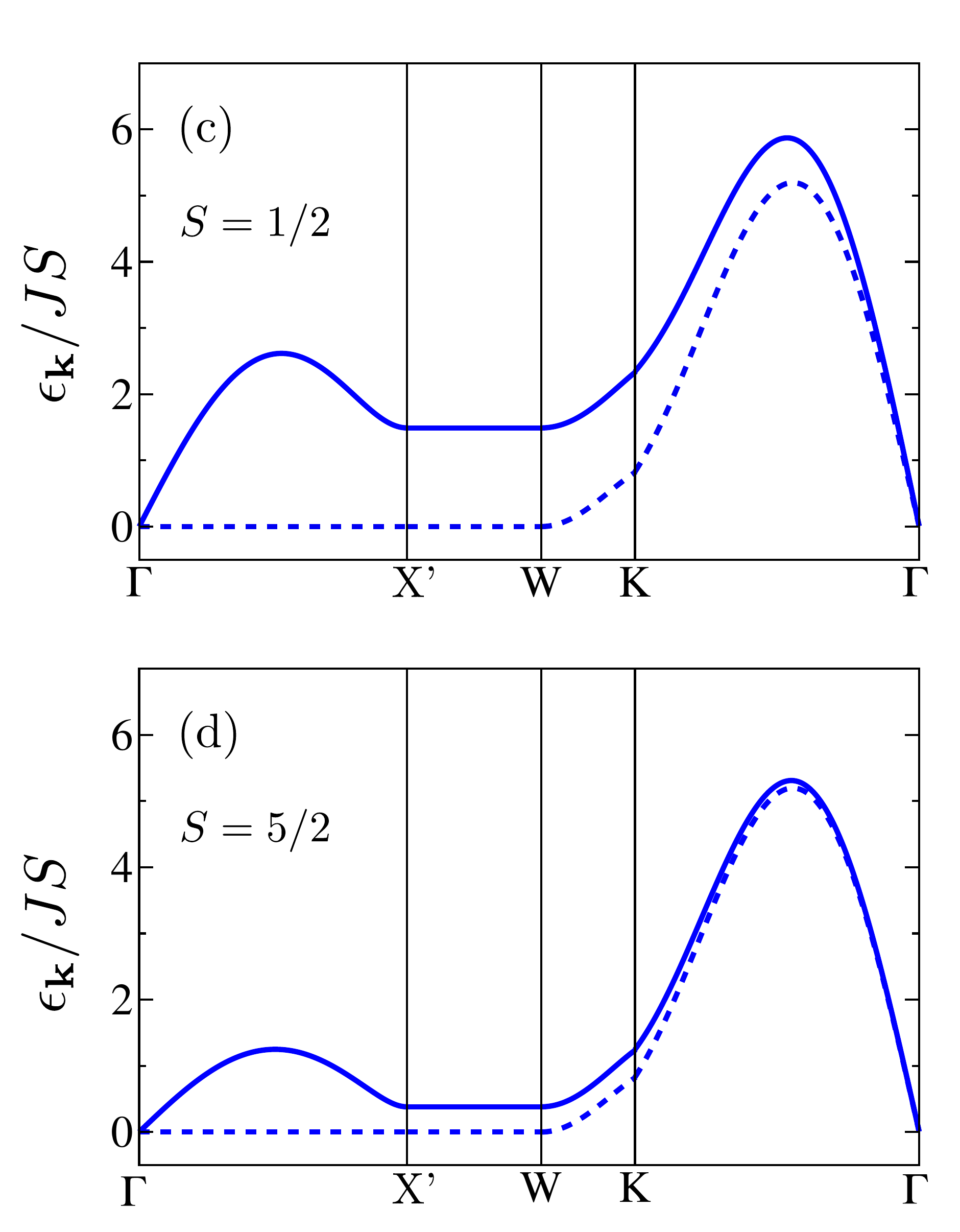}
\caption{Magnon dispersion in the AF1 state.
(a) The Brillouin zone of the fcc lattice with a high-symmetry momentum path. 
(b) False color plots of $\epsilon_{\bf k}$ within the $k_y=2\pi$ plane
computed in the harmonic  theory (upper panel) and in the self-consistent 
approximation for $S=1/2$ (lower panel).
Spin-wave dispersions along the high-symmetry path for (c) $S = 1/2$  and (d) $S=5/2$. 
Full lines show the results of the self-consistent calculations, dashed lines indicate the harmonic
spectra.
}
\label{fig:ek} 
\end{figure}

The collinear AF1 states have a propagation vector at one of the X points in the Brillouin zone, see Fig.~\ref{fig:ek}.
Between degenerate antiferromagnetic domains we choose the state described by ${\bf Q}_1 = (2\pi,0,0)$.
The harmonic spectrum of the AF1 state is obtained from general expressions (\ref{AkBk1}) and (\ref{ek1}) by keeping 
terms $O(JS)$ for $A_{\bf k}$ and $B_{\bf k}$. A zero-energy mode appears
once the harmonic parameters $A_{\bf k},B_{\bf k}$ obey: (i) $A_{\bf k} = |B_{\bf k}|$ or (ii) $A_{\bf k}=B_{\bf k}=0$.
For the fcc antiferromagnet, the pseudo-Goldstone magnons of the first type appear on 
the lines $(0,q,0)$, $(0,q,2\pi)$, and other equivalent directions in the momentum space.
Zero-energy modes of the second type correspond to excitations with the wave vectors $(q,2\pi,0)$ and $(q,0,2\pi)$.
Expanding $A_{\bf k},B_{\bf k}$ in the vicinity of these lines, one can straightforwardly show that the magnon energy vanishes linearly
$\epsilon_{\bf k} \sim k$ for the type-I modes and quadratically $\epsilon_{\bf k} \sim k^2$ for
the type-II modes. The top panel of Fig.~\ref{fig:ek}(b) 
shows the harmonic spectrum of the AF1 state in the plane $k_y=2\pi$. The
zero-energy modes of two types are present as dark blue valleys of different width that
cross at  ${\rm X}' = (0,2\pi,0)$. Note, that ${\rm X}'$ is not the ordering wave vector for the chosen
AF1 state. 

The above classification of pseudo-Goldstone modes can be extended to a general multisublattice case 
beyond the simple  expression (\ref{ek1}) \cite{Rau18}. It is reminiscent of distincting the true Goldstone modes 
for systems with nonconserved (type-I) and conserved (type-II) order parameters, which represent respectively 
the usual  Heisenberg antiferromagnets and ferromagnets \cite{Watanabe12}. Once quantum corrections to 
the spectrum are included within the  $1/S$ expansion, the harmonic $A_{\bf k},B_{\bf k} = O(JS)$ receive 
extra contributions $\delta A_{\bf k},\delta B_{\bf k} = O(J)$. A simple consideration shows that in such a case
an energy of a type-I mode increases as $\Delta_g = O(JS^{1/2})$, whereas a type-II  magnon
acquires a smaller gap $\Delta_g = O(J)$ \cite{Chubukov92,Rau18}.

Figures \ref{fig:ek}(b), \ref{fig:ek}(c), and \ref{fig:ek}(d)  illustrate quantum renormalization of the
magnon spectrum for the AF1 state obtained from the self-consistent spin-wave theory.
The false color maps in Fig.~\ref{fig:ek}(b) compare the harmonic and the renormalized spectrum
in the  $(k_x,2\pi,k_z)$ plane for $S=1/2$. Figs.~\ref{fig:ek}(c) and \ref{fig:ek}(d) show $\epsilon_{\bf k}$  
for $S=1/2$ and 5/2, respectively, along a symmetric path in the Brillouin zone indicated in Fig.~\ref{fig:ek}(a).  
As $S$ increases, the effect of magnon-magnon interaction weakens and, ultimately, the harmonic spectrum 
should be recovered for $S\to\infty$. Such a tendency is illustrated by considering $\epsilon_{\bf k}$ on 
the K--$\Gamma$ segment in Figs.~\ref{fig:ek}(c) and \ref{fig:ek}(d), where magnons have finite energy
already in the harmonic approximation. 

The segments $\Gamma$--${\rm X}'$ and ${\rm X}'$--W 
correspond to the pseudo-Goldstone modes. Magnons along these lines are shifted to finite energies 
by quantum corrections. For the segment ${\rm X}'$--W the energy gap is explicitly given by
\begin{equation}
\Delta_g = 8J(\Delta - m) \ . 
\label{g1}
\end{equation}
Flat magnon dispersion along this line is accidental and a weak modulation of $\epsilon_{\bf k}$
should arise as a result of higher-order scattering processes excluded in the self-consistent approximation.
Note, that the gap obtained by computing the $1/S$ correction to the spectrum has the same form (\ref{g1})
but its value is twice as large as  the self-consistent result, cf.\ Eqs.~(\ref{MF1}) and (\ref{MF10}). 
Comparing results for $S=1/2$ and $S=5/2$ one can also observe different scaling of magnon energies on
the paths  $\Gamma$--X$'$ and X$'$--W, which correspond to type-I and type-II pseudo-Goldstone 
modes, respectively.

% ==============================================================================
% ==============================================================================

Zero-energy modes of the renormalized spectrum correspond to two Goldstone modes at $\Gamma$ and X points. 
The velocity of acoustic magnons is anisotropic  with the two principal values
\begin{equation}
c_\parallel = 4J(S-n+\Delta), \ \ c_\perp = 2J\sqrt{2(S-n+\Delta)(\Delta-m)},
\label{c1}
\end{equation}
where $\parallel$ and $\perp$ are taken with respect to the $\Gamma$--X direction. 
The dispersion along the $\Gamma$--X line is finite in the harmonic approximation, hence,
$c_\parallel= O(JS)$. The  harmonic spectrum has two orthogonal lines
of zero-energy  modes in the $y$--$z$ plane and, as a result, the corresponding
velocity in the renormalized spectrum is generally smaller $c_\perp= O(JS^{1/2})$.

Similar results for the spectrum renormalization  were  also obtained for the AF3 state. 
The corresponding plots are presented in Appendix \ref{app:AF3}. Here we summarize the main qualitative features.
The majority of pseudo-Goldstone modes for the AF3 state belong to the type I. 
If we choose for the ordering wave vector of the AF3 state ${\bf Q}_3 = (2\pi,\pi,0)$, then
the only  type-II pseudo-Goldstone mode exists at ${\rm X}'=(0,2\pi,0)$ and equivalent points.
The expression for the quantum gap at this point 
\begin{equation}
\Delta_g = 8J\sqrt{(\Delta_{aa} - m)(\Delta_{ab}-m)}
\label{g3}
\end{equation}
resembles Eq.~(\ref{g1}) for the AF1 state.
The anisotropic velocity for the acoustic modes in the AF3 state is given by
\begin{eqnarray}
c_\parallel & = & 2J\sqrt{[2(S-n)+\Delta_{aa}+\Delta_{ab}](\Delta_{ab}-m)}\,, 
\nonumber \\
c_\perp & = & J\bigl[4(S-n)+2\Delta_{aa}+2\Delta_{ab}\bigr]^{1/2} \\
& &\mbox{} \times \bigl[2(S-n)+2\Delta_{aa}+\Delta_{ab}-m\bigr]^{1/2}\,.
\nonumber
\label{c2}
\end{eqnarray}
In contrast to the results for the AF1 state (\ref{c1}), two components of the magnon velocity
behave as  $c_\perp=O(JS)$ and one as $c_\parallel=O(JS^{1/2})$.

All the above allows us to conclude that the quantum excitation spectrum  in the AF1 state for the
large-$S$ fcc antiferromagnet is intrinsically softer than in the AF3 state.
This qualitative conclusion has important implication for the low-temperature behavior. In a 3D case,  for $T\ll J$,
an acoustic mode contributes
\begin{equation} 
\Delta F/V = -  \frac{\pi^2}{90}\,\frac{T^4}{c_{\parallel} c_{\perp}^2}\ ,
 \label{dF}
\end{equation}
to the free energy per volume. Since $c_{\parallel} c_{\perp}^2\simeq S^2$ 
and $S^{5/2}$ for AF1 and AF3 states, respectively,  stability of the former state is further
enhanced by the thermal fluctuations. At intermediate temperatures $T\sim \Delta_g$,  magnons with energies
above the quantum gap become also excited. Their density of states 
is obviously larger for the AF1 structure, since the gap (\ref{g1}) is present on the lines, whereas
(\ref{g3}) appears only at separate points in the Brillouin zone. Thus, the conclusion that thermal
fluctuations favor the AF1 state should hold also in the intermediate temperature range.
At higher temperature $T\sim J$, thermal corrections to the spectrum become important and no statement
can be made without further studies. Still, we note that the previous harmonic spin-wave analysis \cite{Schick20} as
well as the classical Monte Carlo simulations \cite{Gvozdikova05} all predict the AF1 state due to the thermal 
order by disorder suggesting this selection to be a universal result for the fcc antiferromagnet.
Hence, the scenario with a finite-temperature transition between the AF3 and AF1 states previously discussed in \cite{Schick20} 
may be realized for small $J_2>0$, which favors at the classical level the AF3 state at $T=0$.

%%%%%%%%%%%%%%%%%%%%%%%%%%%%%%%%%%%%%%%%%%%%%%%%%%%%
\section{Conclusions}
\label{sec:Conclusions}
%%%%%%%%%%%%%%%%%%%%%%%%%%%%%%%%%%%%%%%%%%%%%%%%%%%%

The problem of long-range ordering for the nearest-neighbor fcc antiferromagnet was raised
more than fifty years ago \cite{Anderson50,Luttinger51,Li51,Ziman53,Haar62}. In our work we give
for the first time a comprehensive solution of this problem at $T=0$ using the interacting self-consistent spin-wave
theory and the numerical CCM method. We find that the state selection by quantum fluctuations
proceeds differently for $S<S^*$ and $S>S^*$ corresponding, collinear AF1 and  AF3 states, respectively.
The separation point at $S^*\approx 10$ indicates that magnon-magnon interaction plays a significant
role for all physical values of spin.

 We find good agreement between the two theoretical methods for the ground-state energies and  values of ordered
 moments for the  competing collinear states. In addition, the self-consistent spin-wave calculations provide also
 the magnon spectrum renormalization. The magnon-magnon interaction produces finite quantum gaps for
 the pseudo-Goldstone modes leaving only  two acoustic branches in accordance with the spontaneous 
 breaking of the continuous symmetry in the collinear antiferromagnetic state.
 The normal form of the magnon spectra explains why the self-consistent spin-wave calculations provide
 more accurate numerical predictions in comparison with the harmonic theory and the second-order
 $1/S$ results. Previously the self-consistent spin-wave theory was compared to the numerical results 
for the $J_1$--$J_2$ square-lattice antiferromagnet for $J_2>0.5J_1$ \cite{Singh03}. Our work provides
further evidence that the self-consistent approach has good accuracy also for infinitely degenerate frustrated 
spin models.

The spectacular failure of the harmonic theory  to predict the correct ground state as a result of the quantum 
order by disorder  is related to a close proximity $\Delta E\sim 10^{-3}JS$ of the harmonic  energies for 
the two contenders for the ground state. 
In such a case the higher-order quantum  corrections determined by magnon interaction play an
important role. The example of the nearest-neighbor fcc antiferromagnet is by no means unique.
Another example of a close proximity of the harmonic ground-state energies is realized for the 
kagome antioferromagnet in a wide range of applied magnetic fields $0.5H_s< H< H_s$
\cite{Hassan06}. Elucidating the state selection due to magnon interaction for this model
represent an interesting open theoretical problem.

\section*{Aknowledgments}

We acknowledge  helpful discussions with Y. Iqbal and A. L. Chernyshev.
J.R. thanks the Deutsche Forschungsgemeinschaft for financial support  
(DFG RI 615/25-1). M.E.Z. was supported by ANR, France (Grant No. ANR-18-CE05-0023).

%%%%%%%%%%%%%%%%%%%%%%%%%%%%%%%%%%%%%%%%%%%%%%%%%%%%
\begin{appendix}
%%%%%%%%%%%%%%%%%%%%%%%%%%%%%%%%%%%%%%%%%%%%%%%%%%%%
\section{Self-consistent theory for the AF3 state}
\label{app:AF3}
%%%%%%%%%%%%%%%%%%%%%%%%%%%%%%%%%%%%%%%%%%%%%%%%%%%%

% ==============================================================================
\begin{figure}[tb]
\includegraphics[width=0.85\columnwidth]{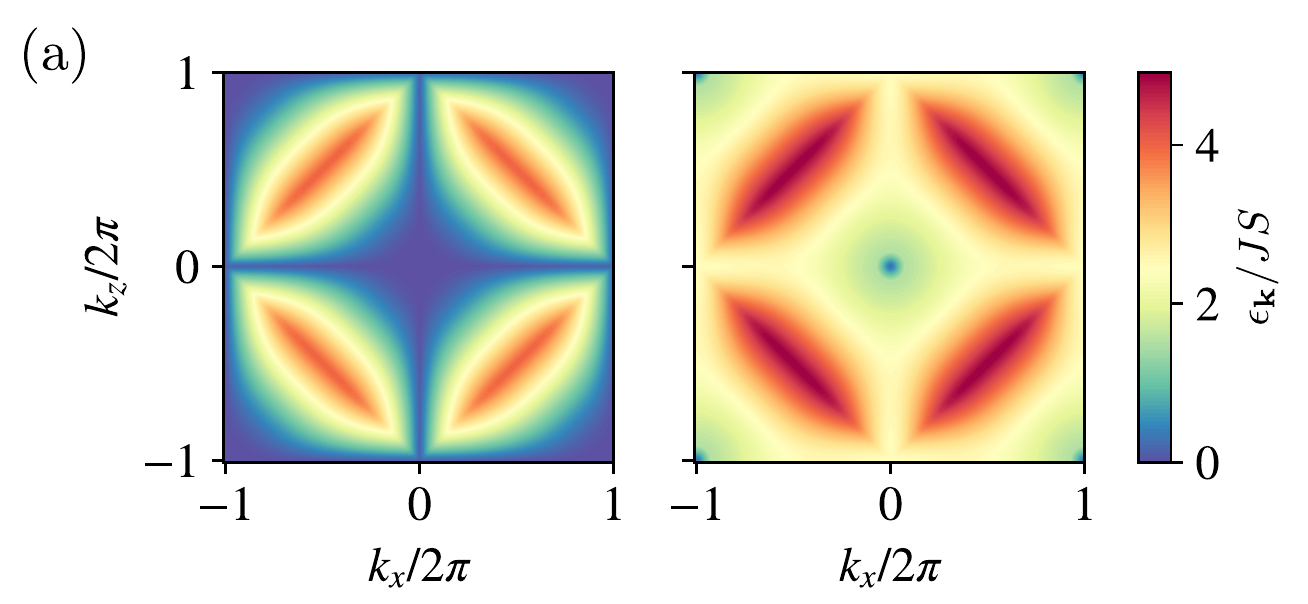}
\vskip 2mm
\includegraphics[width=0.8\columnwidth]{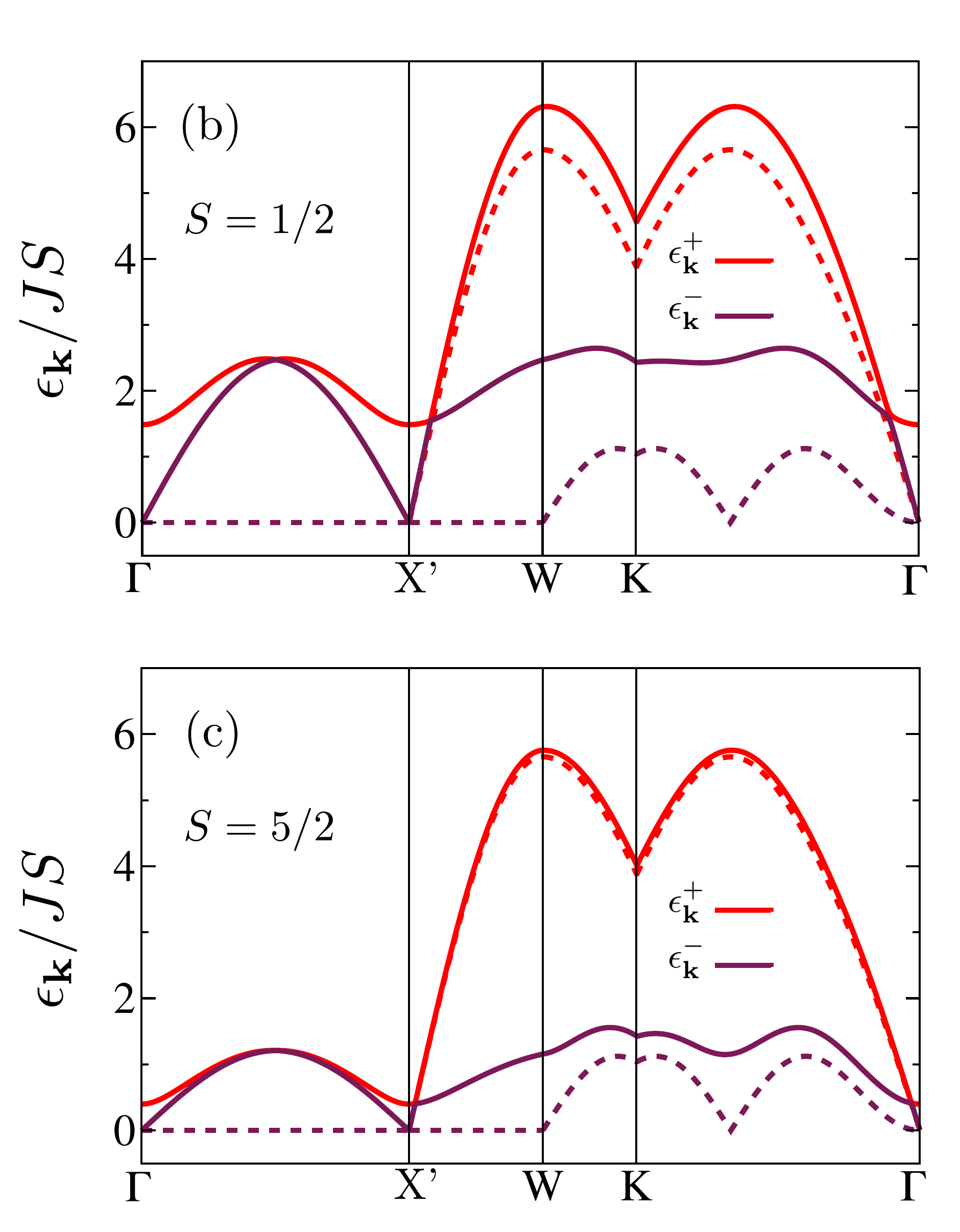}
\caption{Magnon dispersion in the AF3 state.
(a) False color plots of $\epsilon_{\bf k}$ within the $k_y=2\pi$ plane
computed in the harmonic  theory (left panel) and in the self-consistent 
approximation for $S=1/2$ (right panel).
Spin-wave dispersions along the high-symmetry path for (b) $S = 1/2$  and (c) $S=5/2$. 
Full lines show the results of the self-consistent calculations, dashed lines indicate the harmonic
spectra.
}
\label{fig:ek3} 
\end{figure}
% ==============================================================================

In this Appendix we provide additional details on the self-consistent spin-wave calculations 
for the AF3 state. We begin by deriving expressions for the mean-field parameters 
$n$, $m$, $\Delta_{aa}$, and $\Delta_{ab}$
by a method that avoids an explicit use of the Bogolyubov matrix transformation.
The idea consists in adding to the mean-field quadratic Hamiltonian an extra source term linear in a required 
combination of boson operators. For example, to compute  $n = \langle a_i^\dagger a_i \rangle$ we write
\begin{equation}
\hat{\cal H}(h) = \hat{\cal H}^{(2)}_{\rm MF} + h\sum_i a_i^\dagger a_i \ . 
\label{H_field}
\end{equation}
The Hamiltonian $\hat{\cal H}(h)$ can be straightforwardly diagonalized and the magnon spectrum 
is given by the same expression (\ref{ek3}) with a substitution $S_0 \to (S_0+h/4J)$. The  
expectation value in the ground state for the considered combination of boson operators is
obtained as
\begin{equation}
n = \frac{1}{N}\left. \frac{\partial E_{\textrm{g.s.}}(h)}{\partial h}\right|_{h \rightarrow 0} \ ,
\end{equation}
where $N$ is the number of sites. Explicitly, 
\begin{eqnarray}
n + \frac{1}{2}  & = &  \frac{1}{N}\sum_{\bf k, \pm} \frac{2J}{\epsilon^\pm_{\bf k}}  
\biggl\{ 2S_0  \pm  \frac{S_m }{\sqrt{Q_{\bf k}}} \Bigl[2S_0S_m \nonumber  \\
&-&
S_aS_bc_xc_z(\gamma_{\bf k}^2+\gamma_{\bf k}^{*2})\Bigr]
\biggr\}.
\end{eqnarray}
Here the momentum summation is performed over the 
reduced Brillouin zone.  Similar calculation for the other averages yields:
\begin{eqnarray}
m & = &  \frac{1}{N}\sum_{\bf k, \pm} \frac{2J}{\epsilon^\pm_{\bf k}}  
\biggl\{ S_m |\gamma_{\bf k}|^2   \pm  \frac{1}{\sqrt{Q_{\bf k}}} 
\Bigl[2S_0^2S_m |\gamma_{\bf k}|^2
\nonumber \\
&-&
 S_0S_aS_bc_xc_z(\gamma_{\bf k}^2+\gamma_{\bf k}^{*2})
+\frac{1}{2}S_mS_b^2(\gamma_{\bf k}^2-\gamma_{\bf k}^{*2})^2
\Bigr]
\biggr\},
\nonumber \\
\Delta_{aa} & = &  \frac{1}{N}\sum_{\bf k, \pm} \frac{2J}{\epsilon^\pm_{\bf k}}  
\biggl\{ S_a c_x^2c_z^2   \mp  \frac{S_b}{\sqrt{Q_{\bf k}}} 
\Bigl[2S_aS_b c_x^2c_z^2 |\gamma_{\bf k}|^2
\nonumber \\
&-&
 S_0S_mc_xc_z(\gamma_{\bf k}^2+\gamma_{\bf k}^{*2})
\Bigr]
\biggr\},
 \\
\Delta_{ab} & = &  \frac{1}{N}\sum_{\bf k, \pm} \frac{2J}{\epsilon^\pm_{\bf k}}  
\biggl\{ S_b  |\gamma_{\bf k}|^2   \mp  \frac{1}{\sqrt{Q_{\bf k}}} 
\Bigl[2S_a^2S_b c_x^2c_z^2 |\gamma_{\bf k}|^2
\nonumber \\
&-&
 S_0S_mS_a c_xc_z(\gamma_{\bf k}^2+\gamma_{\bf k}^{*2})
 +\frac{1}{2}S_m^2S_b(\gamma_{\bf k}^2-\gamma_{\bf k}^{*2})^2
\Bigr]
\biggr\},
\nonumber 
\end{eqnarray}
The system of four self-consistent equations was solved iteratively.
Values for the four parameters in the AF3 ground state for $S=1/2$ are
$$
n = 0.1297,\  m = 0.0578, \  \Delta_{aa} = 0.1613, \  \Delta_{ab} = 0.1411. 
$$
Again, a significant renormalization is observed in comparison with the harmonic values
$$
n = 0.3663, \ m = 0.0284, \ \Delta_{aa} =0.2943,\ \Delta_{ab} =0.2232.
$$

The magnon dispersion in the AF3 state can be computed using Eq.~(\ref{ek3}) and 
self-consistently obtained parameters $n$, $m$, $\Delta_{aa}$, and $\Delta_{ab}$.
Results for $\epsilon_{\bf k}$ for fcc antiferromagnets with $S=1/2$ and 5/2 that compliment similar plots for the AF1 state in Sec.~\ref{sec:OBDO} are included in Fig.~\ref{fig:ek3}.  We choose a domain of  the AF3 state described by 
${\bf Q}_3 = (2\pi,\pi,0)$ as the propagation vector. In our spin-wave description of the AF3 state in Sec.~\ref{sec:SWT}C
we introduce two (parallel) sublattices. Consequently, we find two different magnon branches and have to consider
accordingly the reduced magnetic Brillouin zone shown in Fig.~\ref{fig:ek3}(a). Still for the plots in Figs.~\ref{fig:ek3}(c)
and \ref{fig:ek3}(d) we use a high-symmetry path in the paramagnetic Brillouin zone. Therefore, 
some wave vectors become equivalent in the reduced Brillouin zone notations. In particular, the ${\rm X}'=(0,2\pi,0)$ point
is equivalent to the  $\Gamma$ point, which explains an extra acoustic mode at ${\rm X}'$ present in Figs.~\ref{fig:ek3}.
%(b) and \ref{fig:ek3}(c).

\begin{table}[bp]
\caption{CCM results for the ground-state energy per site $E_{\rm g.s.}$ and the sublattice magnetization $\langle S\rangle$
in the AF1 and AF3 states with $S=1/2$ and $S=1$.}
\begin{tabular}{|c|cc|cc|} \hline
 $S=1/2^{^{^{^{}}}}$       &\multicolumn{2}{c|}{AF1}       & \multicolumn{2}{c|}{AF3}     \\
                                   &     $E_{\rm g.s.}$        &     $\langle S\rangle$ 	  &    $E_{\rm g.s.}$        &    $\langle S\rangle$      \\  \hline
SUB2-2$^{^{^{^{}}}}$  & $-0.69434$  \; &  0.42655  \;   & $-0.69266$ \; & 0.42830  \;   \\
SUB4-4 	                  & $-0.71708$  \; &  0.39209  \;   & $-0.71557$ \;   & 0.39509  \;  \\
SUB6-6 	                  & $-0.72239$  \; &  0.37529  \;   & $-0.72059$ \;   & 0.37970  \;  \\
SUB8-8 	                  & $-0.72417$  \; &  0.36597  \;   & $-0.72217$ \;   & 0.37162  \;  \\
extra 2-6                    & $-0.72690$  \; &  0.33369  \;   & $-0.72478$ \;    & 0.34244  \;  \\
extra 2-8                    & $-0.72673$  \; &  0.33390  \;   & $-0.72453$ \;    & 0.34346  \;  \\
extra 4-8                    & $-0.72640$  \; &  0.33439  \;   & $-0.72406$ \;    & 0.34585  \;  \\ \hline
\end{tabular}  
\vskip 1mm
\hspace*{0mm}
\begin{tabular}{|c|cc|cc|} \hline
 $S=1^{^{^{^{}}}}$       &\multicolumn{2}{c|}{AF1}       & \multicolumn{2}{c|}{AF3}     \\
                     &     $E_{\rm g.s.}$           &     $M$ 	     &      $E_{\rm g.s.}$        &    $M$         \\  \hline
SUB2-2$^{^{^{^{}}}}$ 	    &   $-2.40691$ \;  &  0.90603    \;     & $-2.40248$  \; &  0.90916  \;  \\
SUB4-4 	    &   $-2.44875$ \;  &  0.86544    \;     & $-2.44598$  \; &  0.86973  \; \\
SUB6-6 	    &   $-2.45734$ \;  &  0.84736    \;     & $-2.45411$  \; &  0.85372 \;  \\
SUB8-8 	    &   $-2.45982$ \;  &  0.83793    \;     & $-$               \; & $-$        \;  \\
extra 2-6      &   $-2.46440$ \;   &  0.80437    \;     & $-2.46063$  \; & 0.81739 \;  \\
extra 2-8      &   $-2.46379$ \;   &  0.80549    \;     & $-$               \; & $-$     \;  \\
extra 4-8      &   $-2.46262$ \;   &  0.80807    \;     & $-$               \; & $-$   \;  \\ \hline
\end{tabular}                              
\label{table}                               
\end{table}

\section{CCM results}
\label{app:CCM}

Extrapolation of numerical results of different  SUB$n$-$n$ approximation schemes to $n\to\infty$ 
was performed using Eq.~(\ref{SUBextra}). For $S=1/2$ we have obtained 
the   series  with $n=2,4,6,8$ for both antiferromagnetic structures. 
Accordingly, it is possible to construct three different extrapolations in the $S=1/2$ case that are
summarized in the Table. A small spread
of final values give an estimate for the error bar on the final result quoted in (\ref{Egs_CCM}).

For the spin-1 model the  $n=2$--8 series was obtained only for the AF1 structure. For the AF3 state with
$S=1$, as well as for all $S>1$ we have to rely on the shorter series $n=2,4,6$, which allows only a single extrapolation
according to Eq.~(\ref{SUBextra}). The numerical results for the $S=1$ case are also included  in the Table.

\end{appendix}


\begin{thebibliography}{99}


%%%%%%%%%%%%%%%  Historical FCC Papers  %%%%%%%%%%%%%%%%%%%%%%

\bibitem{Anderson50}
P. W. Anderson, Phys. Rev. {\bf 79}, 705 (1950).

\bibitem{Luttinger51}
J. M. Luttinger, Phys. Rev. {\bf 81}, 1015 (1951).

\bibitem{Li51}
Y.-Y. Li, Phys. Rev. {\bf 84}, 721 (1951).

\bibitem{Ziman53}
J. M. Ziman, Proc. Phys. Soc. A {\bf 66}, 89 (1953).

\bibitem{Haar62}
D. ter Haar and M. E. Lines, Phil. Trans. R. Soc. London A  {\bf 255}, 1 (1962).

\bibitem{Lines63}
M. E. Lines, Proc. R. Soc. London A {\bf 271}, 105 (1963).

\bibitem{Lines64}
M. E. Lines, Phys. Rev. {\bf 271}, A1336 (1964).

\bibitem{Yamamoto72}
Y. Yamamoto and T. Nagamiya, J. Phys. Soc. Jpn. {\bf 32}, 1248 (1972).

\bibitem{Swendsen73}
R. H. Swendsen, J. Phys. C {\bf 6}, 3763 (1973).

\bibitem{Oguchi85}
T. Oguchi, H. Nishimori, and T. Taguchi, J. Phys. Soc. Jpn. {\bf 54}, 4494 (1985).

\bibitem{Henley87}
C. L. Henley, J. Appl. Phys. {\bf 61}, 3962 (1987).

\bibitem{Diep89}
H. T. Diep and H. Kawamura, Phys. Rev. B {\bf 40}, 7019 (1989).

\bibitem{Heinilaa93}
M. T. Heinil{\"a} and A. S. Oja, Phys. Rev. B {\bf 48}, 16514 (1993).

\bibitem{Yildirim98}
T. Yildirim, A. B. Harris, and E. F. Shender, Phys. Rev. B {\bf 58}, 3144 (1998).

\bibitem{Lefmann01}
K. Lefmann, and C. Rischel, Eur. Phys. J. B {\bf 21}, 313 (2001).

\bibitem{Gvozdikova05}
M. V. Gvozdikova and M. E. Zhitomirsky, JETP Lett. {\bf 81}, 236 (2005).

\bibitem{Ignatenko08}
A. N. Ignatenko, A. A. Katanin, and V. Y. Irkhin, JETP Lett. {\bf 87}, 555 (2008).

\bibitem{Cook15}
A. M. Cook, S. Matern, C. Hickey, A. A. Aczel, and A. Paramekanti,
Phys. Rev. B {\bf 92}, 020417 (2015).

\bibitem{Batalov16}
L. A. Batalov and A. V. Syromyatnikov, J. Magn. Magn. Mater. {\bf 414}, 180 (2016).

\bibitem{Sinkovicz16}
P. Sinkovicz, G. Szirmai, and K. Penc, Phys. Rev. B {\bf 93}, 075137 (2016).

\bibitem{Li17}
F.-Y. Li, Y.-D. Li, Y. Yu, A. Paramekanti, and G. Chen, Phys. Rev. B {\bf 95}, 085132 (2017).

\bibitem{Singh17}
A. Singh, S. Mohapatra, T. Ziman, and T. Chatterji, J.~Appl. Phys. {\bf 121}, 073903 (2017).

\bibitem{Sun18}
N.-N. Sun and H.-Y. Wang, J. Magn. Magn. Mater. {\bf 454}, 176 (2018).

\bibitem{Balla20}
P. Balla, Y. Iqbal, and K. Penc Phys. Rev. Research {\bf 2}, 043278 (2020).

\bibitem{Schick20}
R. Schick, T. Ziman, and M. E. Zhitomirsky, Phys. Rev. B {\bf 102}, 220405 (2020).

\bibitem{Kiese20}
D. Kiese, T. Mueller, Y. Iqbal, R. Thomale, and S. Trebst, preprint
{\tt arXiv:2011.01269}.



%%%%%%%%%%%%%%%%%%%%%%%%%%%%%%%%%%%%%%%%
\bibitem{Wannier50}
G. H. Wannier, Phys. Rev. {\bf 79}, 357 (1950).

%%%%%%%%%%%%%  Experimental works  %%%%%%%%%%%%%%

\bibitem{Seehra88}
M. S. Seehra and T. M. Giebultowicz, Phys. Rev. B {\bf 38}, 11898 (1988).

\bibitem{Matsuura03}
M. Matsuura, Y. Endoh, H. Hiraka, K. Yamada, A. S. Mishchenko, N. Nagaosa, 
and I. V. Solovyev, Phys. Rev. B {\bf 68}, 094409 (2003).

\bibitem{Goodwin07}
A. L. Goodwin, M. T. Dove, M. G. Tucker, and D. A. Keen,
Phys. Rev. B {\bf 75}, 075423 (2007).

\bibitem{Aczel16}
A. A. Aczel, A. M. Cook, T. J. Williams, S. Calder, A. D. Christianson, G.-X. Cao, 
D. Mandrus, Y.-B. Kim, and A. Paramekanti, Phys. Rev. B {\bf 93}, 214426 (2016). 

\bibitem{Chatterji19}
T. Chatterji, L. P. Regnault, S. Ghosh. and A. Singh,
J. Phys.: Condens. Matter  {\bf 31}, 125802 (2019).

\bibitem{Khan19}
N. Khan, D. Prishchenko, Y. Skourski, V. G. Mazurenko, and A. A. Tsirlin,
Phys. Rev. B {\bf 99}, 144425 (2019).

\bibitem{Revelli19}
A. Revelli, C. C. Loo, D. Kiese, P. Becker, T. Frohlich, T. Lorenz, M. Moretti Sala, 
G. Monaco, F. L. Buessen,  J. Attig, M. Hermanns, S. V. Streltsov, D. I. Khomskii, 
J. van den Brink, M. Braden,  P. H. M. van Loosdrecht, S. Trebst, A. Paramekanti, 
and M. Gruninger, Phys. Rev. B {\bf 100}, 085139 (2019).

\bibitem{Bhaskaran21}
L. Bhaskaran, A. N. Ponomaryov, J. Wosnitza, N. Khan, A. A. Tsirlin, M. E. Zhitomirsky, 
and S. A. Zvyagin, Phys. Rev. B {\bf 104}, 184404 (2021).

%%%%%%%%%%%%%%%%%%  Order By Disorder  %%%%%%%%%%%%%%%

\bibitem{Shender82}
E. F. Shender, Zh. Eksp. Teor. Fiz. {\bf 83}, 326 (1982) [Sov. Phys. JETP {\bf 56}, 
178 (1982)].

\bibitem{Henley89}
C. L. Henley, Phys. Rev. Lett. {\bf 62}, 2056 (1989).

\bibitem{Chubukov91}
A. V. Chubukov and D. I. Golosov, 
J. Phys.: Condens. Matter  {\bf 3}, 69 (1991).

\bibitem{Zhitomirsky15}
M. E. Zhitomirsky,  J. Phys.: Conf. Ser. {\bf 592}, 012110 (2015).

%%%%%%%%%%%%%%%%%%%%%%%%%%%%%%%%%%%%%%

\bibitem{Johnston11}
D. C. Johnston, R. J. McQueeney, B. Lake, A. Honecker, M. E. Zhitomirsky, R. Nath, 
Y. Furukawa, V. P. Antropov, and Y. Singh, Phys. Rev. B {\bf 84}, 094445 (2011).

%%%%%%%%%%%%  Self-Consistent SWT  %%%%%%%%%%%%%%%

\bibitem{Takahashi89}
M. Takahashi, Phys. Rev. B {\bf 40}, 2494 (1989).

\bibitem{Xu90}
J. H. Xu and C. S. Ting,  Phys. Rev. B {\bf 42}, 6861 (1990).

\bibitem{Barabanov90}
A. F. Barabanov and O. A. Starykh, JETP Lett. {\bf 51}, 312 (1990).

\bibitem{Bergomi92}
L. Bergomi and T. Jolicoeur, J. de Phys. I {\bf 2}, 371 (1992).

\bibitem{Irkhin92}
V. Y. Irkhin, A. A. Katanin, and M. I. Katsnelson,
J. Phys.: Condens. Matter {\bf 4}, 5227 (1992).

\bibitem{Gochev94}
I. G. Gochev, Phys. Rev. B {\bf 49}, 9594 (1994).

\bibitem{Dotsenko94}
A. V. Dotsenko and O. P. Sushkov, Phys. Rev. B {\bf 50}, 13821 (1994).

\bibitem{Singh03}
R. R. P. Singh, W. Zheng, J. Oitmaa, O. P. Sushkov, and C. J. Hamer,
Phys. Rev. Lett. {\bf 91}, 017201 (2003).

\bibitem{Uhrig09}
G. S. Uhrig, M. Holt, J. Oitmaa, O. P. Sushkov, and R. R. P. Singh,
Phys. Rev. B  {\bf 79}, 092416 (2009).

\bibitem{Takano11}
J. Takano, H. Tsunetsugu, and M. E. Zhitomirsky, 
J. Phys.: Confer. Series {\bf 320}, 012011 (2011).

\bibitem{Werth18}
A. Werth, P. Kopietz, and O. Tsyplyatyev,
Phys. Rev. B {\bf 97}, 180403(R) (2018).

\bibitem{Yamamoto19}
S. Yamamoto and Y. Noriki,
Phys. Rev. B {\bf 99}, 094412 (2019).

%%%%%%%%%%%  Matrix Bogolyubov transform  %%%%%%%%%%%%

\bibitem{Holstein40}
T. Holstein and H. Primakoff, Phys. Rev. {\bf 58}, 1098 (1940).

\bibitem{White65}
R. M. White, M. Sparks, and I. Ortenburger, Phys. Rev. {\bf 139}, A450 (1965).

\bibitem{Colpa78}
J. H. P. Colpa, Physica A {\bf 93}, 327 (1978).


%%%%%%%%%%%  Coupled cluster method  %%%%%%%%%%%%%%%


\bibitem{zeng98}
C.~Zeng, D. J. J.~Farnell, and R. F.~Bishop,
J. Stat. Phys. {\bf 90}, 327 (1998).

\bibitem{darradi08}
R. Darradi, O. Derzhko, R. Zinke, J. Schulenburg, S. E. Kr\"uger and J. Richter,
Phys. Rev. B {\bf 78}, 214415 (2008).

\bibitem{farnell09}
D. J. J.~Farnell, R.~Zinke,  J.~Schulenburg, and J.~Richter,
J. Phys.: Cond. Matter {\bf 21}, 406002 (2009).

\bibitem{kagome_general_s_2011} 
O. G\"otze, D. J. J. Farnell, R. F. Bishop, P. H. Y. Li, and J. Richter, 
Phys. Rev. B {\bf 84}, 224428 (2011).

\bibitem{archi2014}  
D. J. J. Farnell, O. G\"otze,  J. Richter, R. F. Bishop,  and
P. H. Y.~Li, Phys. Rev. B {\bf 89}, 184407 (2014).

\bibitem{sc-bcc-2015}  
D. J. J. Farnell, O. G\"otze, and J. Richter, Phys. Rev. B {\bf  93}, 235123 (2016).

\bibitem{jiang2015}
J.-J. Jiang, F. Tang, and C. H. Yang, 
J. Phys. Soc. Jpn.  {\bf 84}, 124710 (2015).

\bibitem{archi2018}
D. J. J. Farnell, O. G\"otze, J. Schulenburg, R. Zinke, R. F. Bishop, 
P. H. Y. Li, Phys. Rev. B {\bf 98}, 224402 (2018)

\bibitem{bilayer2019}  
R.F.~Bishop, P.H.Y.~Li, O.~G\"{o}tze, and J. Richter,
Phys. Rev. B {\bf 100}, 024401 (2019).

\bibitem{jian2021}
J.-J. Jiang, J. Magn. Magn. Mat. {\bf 539}, 168392 (2021).

\bibitem{SUBn-n} For $s=1/2$  typically the notation LSUB$n$ is
used. Note that for $s=1/2$ the LSUB$n$ scheme is identical 
to the SUB$n$-$n$ scheme.    

\bibitem{cccm} 
For the numerical calculation we use the program package `The
crystallographic CCM' (D. J. J. Farnell and J. Schulenburg).

%%%%%%%%%%%%%  Quantum Gaps  %%%%%%%%%%%%%%%

\bibitem{Belorizky80}
E. Belorizky, R. Casalegno, and J. J. Niez,
Phys. Stat. Sol. (b) {\bf 102}, 365 (1980).

\bibitem{Chubukov92}
A. V. Chubukov and T. Jolicoeur, Phys. Rev. B {\bf 46}, 11137 (1992).

\bibitem{Rau18}
J. G. Rau, P. A. McClarty, and R. Moessner, Phys. Rev. Lett. {\bf 121}, 237201 (2018).

\bibitem{Watanabe12}
H. Watanabe and H. Murayama, Phys. Rev. Lett. {\bf 108}, 251602 (2012).

\bibitem{Hassan06}
S. R. Hassan and R. Moessner, Phys. Rev. B {\bf 73}, 094443 (2006).

\end{thebibliography}
\end{document}